\RequirePackage{snapshot}
\documentclass[12pt]{article}

\usepackage[T1]{fontenc}
\usepackage[utf8]{inputenc}

\usepackage
[a4paper,tmargin=3truecm,bmargin=3truecm,rmargin=2.5truecm,lmargin=2.5truecm,twoside,verbose=tr
ue]{geometry}
\usepackage{amsmath,amssymb,amsthm,latexsym}

\usepackage{stmaryrd,wasysym,upgreek,mathrsfs,dsfont}
\usepackage[english]{babel}
\usepackage{graphicx,color}
\usepackage{slashed,subfig}

\theoremstyle{plain}
\newtheorem{thm}{Theorem}[section]
\newtheorem{lemma}[thm]{Lemma}
\newtheorem{prop}[thm]{Proposition}
\newtheorem{cor}[thm]{Corollary}

\theoremstyle{plain}
\newtheorem{defn}{Definition}[section]
\newtheorem*{rem}{Remark}

\newcommand{\beqa}{\begin{eqnarray}}
\newcommand{\eeqa}{\end{eqnarray}}
\newcommand{\bea}{\begin{eqnarray}}
\newcommand{\eea}{\end{eqnarray}}

\renewcommand{\mathbf}{\boldsymbol}

\newcommand{\encv}{{non-commutative}}

\newcommand{\les}{\leqslant}
\newcommand{\ges}{\geqslant}

\newcommand{\N}{\mathbb{N}}

\newcommand{\R}{\mathbb{R}}
\newcommand{\C}{\mathbb{C}}

\newcommand{\bbbone}{{\mathds{1}}}

\newcommand{\lnat}{\llbracket} 
\newcommand{\rnat}{\rrbracket} 

\newcommand{\defi}{\stackrel{\text{\tiny def}}{=}}
\newcommand{\wed}{\wedge}

\def\lbt{\left(}
  \def\rbt{\right)}
\def\labs{\left |}
  \def\rabs{\right |}

\def\lb{\left \{}
  \def\rb{\right \}}
\def\lsb{\left [}
  \def\rsb{\right ]}

\DeclareMathOperator*{\Tr}{Tr}

\DeclareMathOperator*{\seq}{\simeq}

\newcommand {\tqs}{\mathrel{:}}

\newcommand{\GN}{\text{GN}^{2}_{\Theta}}

\def\xt{\widetilde{x}}

\def\Ot{\widetilde{\Omega}}

\def\ps{\slashed{p}}

\def\xs{\slashed{x}}
\def\ys{\slashed{y}}

\def\xts{\slashed{\widetilde{x}}}
\def\yts{\slashed{\widetilde{y}}}

\def\psib{\bar{\psi}}

\newcommand\scF{{\mathscr F}}

\newcommand\cA{{\mathcal A}}

\newcommand\cI{{\mathcal I}}

\newcommand\cM{{\mathcal M}}

\newcommand\cO{{\mathcal O}}

\newcommand\cS{{\mathcal S}}

\newcommand{\noi}{\noindent}

\numberwithin{equation}{section}

\begin{document}

\title{Renormalisation\\ of non-commutative field theories}
\author{Vincent Rivasseau$^{a}$\footnote{This review follows lectures given by V.R.
at the workshop  ``Renormalisation et th\'eories de Galois'',  Luminy, March 2006.}\\
Fabien Vignes-Tourneret$^{b}$}

\maketitle
\vspace*{-1cm}
\begin{center}
\textit{$^a$Laboratoire de Physique Théorique, Bât.\ 210\\
    Université Paris XI,  F-91405 Orsay Cedex, France\\
    e-mail: \texttt{rivass@th.u-psud.fr}}\\
  \textit{$^b$IHÉS, Le Bois-Marie, 35 route de Chartres, F-91440 Bures-sur-Yvette, France\\
    e-mail: \texttt{vignes@ihes.fr}}
 \end{center}%

\begin{abstract}
The first renormalisable quantum field theories on non-commutative
space have been found recently. We review this rapidly growing subject. 
\end{abstract}

\tableofcontents

\section{Introduction}

General relativity and ordinary differential geometry
should be replaced by non-commutative geometry
at some point between the currently accessible 
energies of about 1 - 10 Tev (after starting the Large Hadron Collider (LHC) at CERN) 
and the Planck scale, which is $10^{15}$ times higher,
where space-time and gravity should be quantized. 

This could occur either at the Planck scale or below. Quantum field theory
on a non-commutative space-time (NCQF) 
could very well be an intermediate
theory relevant for physics at energies
between the LHC and the Planck scale.
It certainly looks intermediate in structure
between ordinary quantum field theory on commutative ${\mathbb R^4}$ 
and string theory, the current 
leading candidate for a more fundamental theory including quantized gravity. 
NCQFT in fact appears as an effective model for certain limits of string theory
\cite{a.connes98noncom,Seiberg1999vs}.

In joint work with R. Gurau, J. Magnen and F. Vignes-Tourneret \cite{xphi4-05},
using direct space methods, we provided recently a new proof that the Grosse-Wulkenhaar
scalar $\Phi^4_4$ theory on the Moyal space ${\mathbb R}^4$
is renormalisable to all orders in perturbation theory. 

The Grosse-Wulkenhaar breakthrough \cite{GrWu03-1,c}
was to realize that the right propagator in non-commutative field theory 
is not the ordinary commutative propagator, but has to be modified to obey Langmann-Szabo duality 
\cite{LaSz,c}. Grosse and Wulkenhaar were able to compute the corresponding propagator in the so 
called ``matrix base'' which transforms the Moyal product 
into a matrix product. This is a real {\it tour de force}! They use this representation to prove 
perturbative renormalisability of the theory up to some estimates 
which were finally proven in \cite{Rivasseau2005bh}.

Our direct space method builds upon the previous works of Filk and Chepelev-Roiban
\cite{Filk1996dm,Chepelev2000hm}. These works however remained inconclusive 
\cite{CheRoi}, since these authors used the right interaction but not the right 
propagator, hence the problem of ultraviolet/infrared mixing prevented them 
from obtaining a finite renormalised perturbation series.

We also extend the Grosse-Wulkenhaar results to more
general models with covariant derivatives in a fixed magnetic field \cite{Langmann2003if}.
Our proof relies on a multiscale analysis analogous to \cite{Rivasseau2005bh}
but in direct space.

Non-commutative field theories (for a general review see \cite{DouNe}) deserve a thorough 
and systematic investigation, not only because they may be relevant for physics beyond the standard 
model,
but also (although this is often less emphasized) because they
can describe effective physics in our ordinary standard world but
with non-local interactions.

In this case there is an interesting reversal of
the initial Grosse-Wulkenhaar problematic. In the
$\Phi^4_4$ theory on the Moyal space ${\mathbb R}^4$,
the vertex is sort of God-given by the Moyal structure, and it is LS invariant. The challenge was
to overcome uv/ir mixing and to find the right propagator 
which makes the theory renormalisable. This propagator turned out to
have LS duality. The harmonic potential introduced by Grosse and Wulkenhaar
can be interpreted as a piece of covariant derivatives in a constant magnetic field. 

Now to explain the (fractional) quantum Hall effect, which is a bulk effect
whose understanding requires electron interactions, we can almost invert this logic. The propagator
is known since it corresponds to non-relativistic electrons in two dimensions
in a constant magnetic field. It has LS duality. But the interaction is unclear, and cannot be local
since at strong magnetic field the spins should align with the magnetic field, hence by Pauli
principle local interactions among electrons in the first Landau level should vanish.

We can argue that among all possible non-local interactions, a few renormalisation
group steps should select the only ones which form a renormalisable theory with the corresponding
propagator. In the commutative case (i.e. zero magnetic field)
local interactions such as those of the Hubbard model are just 
renormalisable in any dimension because of the extended nature of the Fermi-surface singularity.
Since the non-commutative electron propagator (i.e. in non zero magnetic field)
looks very similar to the Grosse-Wulkenhaar propagator (it is in fact a 
generalization of the Langmann-Szabo-Zarembo propagator)
we can conjecture that the renormalisable interaction corresponding to this propagator
should be given by a Moyal product. That's why we hope
that non-commutative field theory is the correct framework for a 
microscopic {\it ab initio} understanding 
of the fractional quantum Hall effect which is currently lacking.

Even for regular commutative field theory such as non-Abelian gauge theory,
the strong coupling or non-perturbative regimes may be studied fruitfully through
their non-commutative (i.e. non local) counterparts. 
This point of view is forcefully suggested in \cite{Seiberg1999vs}, where a mapping is proposed
between ordinary and non-commutative gauge fields which do not preserve the gauge groups
but preserve the gauge equivalent classes.
We can at least remark that the effective physics of confinement should be
governed by a non-local interaction, as is the case in effective strings or bags models.

In other words we propose to base physics upon the renormalisability principle, more than any other 
axiom. Renormalisability means genericity; only renormalisable interactions survive
a few RG steps, hence only them should be used to describe generic effective physics of any kind.
The search for renormalisabilty could be the powerful principle on which to orient ourselves 
in the jungle of all possible non-local interactions.

Renormalisability has also attracted considerable interest in the recent years
as a pure mathematical structure. The work of Kreimer and Connes \cite{Kreimer:1997dp,Connes:2000uq,Connes:2001kx}
recasts the recursive BPHZ forest formula of
perturbative renormalisation in a nice Hopf algebra structure. 
The renormalisation group ambiguity reminds mathematicians of
the Galois group ambiguity for roots of algebraic equations.
Finding new renormalisable theories may therefore be 
important for the future of pure mathematics as well as for physics.
That was forcefully argued during the Luminy workshop ``Renormalisation and Galois Theory''.
Main open conjectures in pure mathematics such as 
the Riemann hypothesis \cite{Connes:2004xb,Leichtnam2006a}
or the Jacobian conjecture \cite{Abdesselam:2002cy} may benefit from the
quantum field theory and renormalisation group approach.

Considering that most of the Connes-Kreimer works uses
dimensional regularization and the minimal dimensional
renormalisation scheme, it is interesting to develop the parametric representation which generalize
Schwinger's parametric representation of Feynman amplitudes
to the non commutative context. It involves
hyperbolic generalizations of the ordinary topological polynomials, which mathematicians
call Kirchoff polynomials, and physicist call Symanzik polynomials in the quantum field 
theory context \cite{gurauhypersyman}.
We plan also to work out the corresponding regularization and minimal dimensional
renormalisation scheme and to recast it in a Hopf algebra structure.
The corresponding structures seem richer than in ordinary field theory since
they involve ribbon graphs and invariants which contain information about the
genus of the surface on which these graphs live.

A critical goal to enlarge the class of renormalisable non-commutative field theories and
to attack the Quantum Hall effect problem is to extend the results of Grosse-Wulkenhaar
to Fermionic theories.
The simplest theory, the two-dimensional Gross-Neveu model can be shown 
renormalisable to all orders in their Langmann-Szabo covariant versions, using either the matrix basis
\cite{toolbox05}
or the direct space version developed here \cite{RenNCGN05}. However the $x$-space version seems 
the most 
promising for a complete non-perturbative construction, using Pauli's principle to controll the
apparent (fake) divergences of perturbation theory. 

In the case of $\phi^4_4$, recall that although the commutative version is until now fatally flawed 
due to the famous Landau ghost, there is hope that the non-commutative field theory treated 
at the perturbative level in this paper may also exist at the constructive level.
Indeed a non trivial fixed point of the renormalization group develops at
high energy, where the Grosse-Wulkenhaar parameter $\Omega$ tends to 1,
so that Langmann-Szabo duality become exact, and the beta function vanishes. 
This scenario has been checked explicitly to all orders of perturbation theory
\cite{GrWu04-2,DisertoriRivasseau2006,beta2-06}. 
This was done using the matrix version of the theory; again
an $x$-space version of renormalisation might be better
for a future rigorous non-perturbative investigation of this fixed point
and a full constructive version of the model.

Finally let us conclude this short introduction by reminding that a very important and difficult goal
is to also extend the Grosse-Wulkenhaar breakthrough to gauge theories.

\subsection{The Quantum Hall effect}
One considers free electrons:
 $H_{0}=\frac{1}{2m}({\mathbf p}+e{\mathbf A})^{2}=\frac{\pi^{2}}{2m}$
where ${\mathbf p}=m\dot{\bf r}-e{\bf A}$ is the canonical conjugate of ${\bf
  r}$.

The moment and position ${\bf p}$ and ${\bf r}$ have
commutators
\begin{eqnarray}
[p_i,p_j]=0,\ \ [r_i,r_j]=0,\ \ [p_i,r_j]=\imath\hbar\delta_{ij} .
\end{eqnarray}

The moments $\mbox{\boldmath $\pi$}=m\dot{{\bf r}}={\bf p}+e{\bf A}$ 
have commutators
\begin{eqnarray}
  [\pi_i,\pi_j]=-\imath\hbar\epsilon_{ij}eB,\ \ [r_i,r_j]=0,\ \
  [\pi_i,r_j]=\imath\hbar\delta_{ij} .
\end{eqnarray}

One can also introduce coordinates $R_{x}$, $R_{y}$
corresponding to the centers of the classical trajectories
\begin{eqnarray}
  R_x=x-\frac{1}{e B}\pi_y,\ \ R_y=y+\frac{1}{e B} \pi_x
\end{eqnarray}
which {\it do not commute}:
\begin{eqnarray}
{ [R_i,R_j]=\imath\hbar\epsilon_{ij}\frac{1}{e B}},\ \ [\pi_i,R_j]=0.
 \end{eqnarray}
 This means that there exist Heisenberg-like relations
between quantum positions.

\subsection{String Theory in background field}

One considers the string action in a generalized 
background
\begin{eqnarray}
S&=&\frac{1}{4\pi \alpha ' }\int_{\Sigma} 
( g_{\mu\nu}\partial_a X^{\mu} \partial^a X^{\nu}
{-2\pi i \alpha '  B_{\mu \nu} \epsilon^{ab} \partial_a X^{\mu} \partial_b X^{\nu}} )
\\
&=&\frac{1}{4\pi \alpha ' }\int_{\Sigma} 
 g_{\mu\nu}\partial_a X^{\mu} \partial^a X^{\nu}- \frac{i}{2}\int_{\partial \Sigma}
B_{\mu\nu} X^{\mu} \partial_t X^{\nu} ,
\end{eqnarray}
where $\Sigma $ is the string worldsheet, $\partial_t$ is a tangential derivative
along the worldsheet boundary $\partial \Sigma$ and
$B_{\mu \nu}$ is an antisymmetric background tensor.
The equations of motion determine the boundary conditions: 
\begin{eqnarray}
g_{\mu\nu}\partial_n X^{\mu} 
+ 2\pi i \alpha '  B_{\mu \nu} \partial_t X^{\mu} \vert_{\partial \Sigma} = 0 .
\end{eqnarray}

Boundary conditions for coordinates can be Neumann ($B \to 0$) or Dirichlet 
($g \to 0$, corresponding to branes).

After conformal mapping of the string worldsheet onto the upper half-plane,
the string propagator in background field is
\begin{eqnarray}
< X^{\mu}(z) X^{\nu}(z') > &=& - \alpha ' \Big[ g^{\mu\nu} ( \log \vert z - z' \vert
\log \vert z - \bar z' \vert ) \nonumber \\
&&+ G^{\mu\nu} \log \vert z - \bar z' \vert^2  
+ \theta ^{\mu\nu} \log \frac{\vert z - \bar z' \vert}{\vert \bar z - z' \vert} 
+ \text{const}  \Big] \; .
\end{eqnarray}
for some constant symmetric and antisymmetric tensors $G$ and $\theta$.

Evaluated at boundary points on the worldsheet, this propagator is
\begin{eqnarray}
< X^{\mu}(\tau) X^{\nu}(\tau ') > = - \alpha ' G^{\mu\nu} \log (\tau  - \tau ' ) ^2
+ \frac{i}{2}\theta ^{\mu\nu} \epsilon (\tau - \tau ' ) \; ,
\end{eqnarray}
where the $\theta$ term simply
comes from the discontinuity of the logarithm across its cut.
Interpreting $\tau $ as time, one finds
\begin{eqnarray}
[ X^{\mu} , X^{\nu}]= i \theta^{\mu \nu} ,
\end{eqnarray}
which means that string coordinates lie in a non-commutative Moyal space
with parameter $\theta$.

There is an equivalent argument inspired by $M$ theory: a rotation sandwiched between two 
$T$ dualities generates the same constant commutator for string coordinates.

\section{Non-commutative field theory}

\subsection{Field theory on Moyal space}
\label{sec:new-divergences}

The recent progresses concerning the renormalisation of non-commutative field theory have been 
obtained on a very simple non-commutative space namely the Moyal space. From the point of view of 
quantum field theory, it is certainly the most studied space. Let us start with its precise definition.

\subsubsection{The Moyal space \texorpdfstring{${\mathbb R}^{D}_{\theta}$}{}}

Let us define $E=\lb x^{\mu},\,\mu\in\lnat 1,D\rnat\rb$ and $\C\langle E\rangle$ the free algebra 
generated by $E$. Let $\Theta$ a $D\times D$ non-degenerate skew-symmetric matrix (wich requires 
$D$ even) and $I$ the ideal of $\C\langle E\rangle$ generated by the elements $x^{\mu}x^{\nu}-x^{\nu}
x^{\mu}-\imath\Theta^{\mu\nu}$. The Moyal algebra $\cA_{\Theta}$ is the quotient $\C\langle E\rangle/I$. 
Each element in $\cA_{\Theta}$ is a formal power series in the $x^{\mu}$'s for which the relation $\lsb x^
{\mu},x^{\nu}\rsb=\imath\Theta^{\mu\nu}$ holds.

Usually, one puts the matrix $\Theta$ into its canonical form :
\begin{eqnarray}
  \Theta= 
  \begin{pmatrix}
    \begin{matrix} 0 &\theta_{1} \\ 
      \hspace{-.5em} -\theta_{1}&0
    \end{matrix}    &&     (0)
    \\ 
    &\ddots&\\
    (0)&&
    \begin{matrix}0&\theta_{D/2}\\
      \hspace{-.5em}-\theta_{D/2}&0
    \end{matrix}
  \end{pmatrix}.\label{eq:Thetamatrixbase}
\end{eqnarray}
Sometimes one even set $\theta=\theta_{1}=\dotsm =\theta_{D/2}$. The preceeding algebraic definition 
whereas short and precise may be too abstract to perform real computations. One then needs a more 
analytical definition. A representation of the algebra $\cA_{\Theta}$ is given by some set of functions on 
$\R^{d}$ equipped with a non-commutative product: the \emph{Groenwald-Moyal} product. What follows 
is based on \cite{Gracia-Bondia1987kw}.

\paragraph{The Algebra $\cA_{\Theta}$}
\label{sec:lalgebre-ca_theta}

The Moyal algebra $\cA_{\Theta}$ is the linear space of smooth and rapidly decreasing functions $\cS
(\R^{D})$ equipped with the \encv{} product defined by: $\forall f,g\in\cS_{D}\defi\cS(\R^{D})$,
\begin{align}
  (f\star_{\Theta} g)(x)=&\int_{\R^D} \frac{d^{D}k}{(2\pi)^{D}}d^{D}y\, f(x+{\textstyle\frac 12}\Theta\cdot
  k)g(x+y)e^{\imath k\cdot y}\\
  =&\frac{1}{\pi^{D}\labs\det\Theta\rabs}\int_{\R^D} d^{D}yd^{D}z\,f(x+y)
  g(x+z)e^{-2\imath y\Theta^{-1}z}\; .
  \label{eq:moyal-def}
\end{align}
This algebra may be considered as the  ``functions on the Moyal space $\R^{D}_{\theta}$''. In the 
following we will write $f\star g$ instead of $f\star_{\Theta}g$ and use : $\forall f,g\in\cS_{D}$, $\forall j\in
\lnat 1,2N\rnat$,
\begin{align}
  (\scF f)(x)=&\int f(t)e^{-\imath tx}dt  
\end{align}
for the Fourier transform and
\begin{align}
  (f\diamond g)(x)=&\int f(x-t)g(t)e^{2\imath x\Theta^{-1}t}dt  
\end{align}
for the twisted convolution. As on $\R^{D}$, the Fourier transform exchange product and convolution:
\begin{align}
    \scF(f\star g)=&\scF(f)\diamond\scF(g)\label{eq:prodtoconv}\\
    \scF(f\diamond g)=&\scF(f)\star\scF(g)\label{eq:convtoprod}.
\end{align}
One also shows that the Moyal product and the twisted convolution are \textbf{associative}:
\begin{align}
  ((f\diamond g)\diamond h)(x)=&\int f(x-t-s)g(s)h(t)e^{2\imath(x\Theta^{-1}t+(x-t)\Theta^{-1}s)}ds\,dt\\
  =&\int f(u-v)g(v-t)h(t)e^{2\imath(x\Theta^{-1}v-t\Theta^{-1}v)}dt\,dv\notag\\
  =&(f\diamond(g\diamond h))(x).
\end{align}
Using \eqref{eq:convtoprod}, we show the associativity of the $\star$-produit. The complex conjugation 
is \textbf{involutive} in $\cA_{\Theta}$
\begin{align}
  \overline{f\star_{\Theta}g}=&\bar{g}\star_{\Theta}\bar{f}.\label{eq:Moyal-involution}  
\end{align}
One also have
\begin{align}
  f\star_{\Theta}g=&g\star_{-\Theta}f.\label{eq:Moyal-commutation}  
\end{align}
\begin{prop}[Trace]\label{prop:trace}
  For all $f,g\in\cS_{D}$,
  \begin{align}
    \int dx\,(f\star g)(x)=&\int dx\,f(x)g(x)=\int dx\,(g\star f)(x)\; .\label{eq:Moyal-trace}
  \end{align}
\end{prop}
\begin{proof}
  \begin{align}
    \int dx\,(f\star g)(x)=&\scF(f\star g)(0)=(\scF f\diamond\scF g)(0)\\
    =&\int\scF f(-t)\scF g(t)dt=(\scF f\ast\scF g)(0)=\scF(fg)(0)\notag\\
    =&\int f(x)g(x)dx\notag
  \end{align}
where $\ast$ is the ordinary convolution.
\end{proof}

In the following sections, we will need lemma \ref{lem:Moyal-prods} to compute the interaction terms for 
the $\Phi^{4}_{4}$ and Gross-Neveu models. We write $x\wed y\defi 2x\Theta^{-1}y$.
\begin{lemma}\label{lem:Moyal-prods}
  For all $j\in\lnat 1,2n+1\rnat$, let $f_{j}\in\cA_{\Theta}$. Then
  \begin{align}
    \lbt f_{1}\star_{\Theta}
    \dotsb\star_{\Theta}f_{2n}\rbt(x)=&\frac{1}{\pi^{2D}\det^{2}\Theta}\int\prod_{j=1}^{2n}
    dx_{j}f_{j}(x_{j})\,e^{-\imath
      x\wed\sum_{i=1}^{2n}(-1)^{i+1}x_{i}}\,e^{-\imath\varphi_{2n}},\\
    \lbt f_{1}
    \star_{\Theta}\dotsb\star_{\Theta}f_{2n+1}\rbt(x)=&\frac{1}{\pi^{D}\det\Theta}\int\prod_{j=1}^{2n+1}
    dx_{j}f_{j}(x_{j})\,\delta\Big(x-\sum_{i=1}^{2n+1}(-1)^{i+1}x_{i}\Big)\,e^{-\imath\varphi_{2n+1}},\\
  \forall p\in\N,\,\varphi_{p}=&\sum_{i<j=1}^{p}(-1)^{i+j+1}x_{i}\wed x_{j}.
  \end{align}
\end{lemma}
\begin{cor}\label{cor:int-Moyal}
  For all $j\in\lnat 1,2n+1\rnat$, let $f_{j}\in\cA_{\Theta}$. Then
  \begin{align}
    \int dx\,\lbt
    f_{1}\star_{\Theta}\dotsb\star_{\Theta}f_{2n}\rbt(x)=&\frac{1}{\pi^{D}\det\Theta}
    \int\prod_{j=1}^{2n}
    dx_{j}f_{j}(x_{j})\,\,\delta
    \Big(\sum_{i=1}^{2n}(-1)^{i+1}x_{i}\Big)e^{-\imath\varphi_{2n}},\label{eq:int-Moyal-even}\\
    \int dx\,\lbt f_{1}\star_{\Theta}
    \dotsb\star_{\Theta}f_{2n+1}\rbt(x)=&\frac{1}{\pi^{D}\det\Theta}\int\prod_{j=1}^{2n+1}
    dx_{j}f_{j}(x_{j})\,e^{-\imath\varphi_{2n+1}},\\
    \forall p\in\N,\,\varphi_{p}=&\sum_{i<j=1}^{p}(-1)^{i+j+1}x_{i}\wed x_{j}.
  \end{align}
\end{cor}

The cyclicity of the product, inherited from proposition \ref{prop:trace} implies: $\forall f,g,h\in\cS_{D}$,
\begin{align}
  \langle f\star g,h\rangle=&\langle f,g\star h\rangle=\langle g,h\star f\rangle
\end{align}
and allows to extend the Moyal algebra by duality into an algebra of tempered distributions.

\paragraph{Extension by Duality}
\label{sec:extens-par-dual}

Let us first consider the product of a tempered distribution with a Schwartz-class function. Let $T\in\cS'_
{D}$ and $h\in\cS_{D}$. We define $\langle T,h\rangle\defi T(h)$ and $\langle T^\ast,h\rangle =\overline
{\langle T,\overline{h}\rangle}$.
\begin{defn}\label{defn:Tf}
  Let $T\in\cS'_{D}$, $f,h\in\cS_{D}$, we define $T\star f$ and $f\star T$ by
  \begin{align}
    \langle T\star f,h\rangle=&\langle T,f\star h\rangle,\\
    \langle f\star T,h\rangle=&\langle T,h\star f\rangle.
  \end{align}
\end{defn}
For example, the identity $\bbbone$ as an element of $\cS'_{D}$ is the unity for the $\star$-produit: $
\forall f,h\in\cS_{D}$,
\begin{align}
  \langle\bbbone\star f,h\rangle=&\langle\bbbone,f\star h\rangle\\
  =&\int(f\star h)(x)dx=\int f(x)h(x)dx\notag\\
  =&\langle f,h\rangle.\notag
\end{align}
We are now ready to define the linear space $\cM$ as the intersection of two sub-spaces $\cM_{L}$ and 
$\cM_{R}$ of $\cS'_{D}$.
\begin{defn}[Multipliers algebra]\label{defn:M}
  \begin{align}
    \cM_{L}=&\lb S\in\cS'_{D}\tqs\forall f\in\cS_{D},\,S\star f\in\cS_{D}\rb,\\
    \cM_{R}=&\lb R\in\cS'_{D}\tqs\forall f\in\cS_{D},\,f\star R\in\cS_{D}\rb,\\
    \cM=&\cM_{L}\cap\cM_{R}.
  \end{align}
\end{defn}
One can show that $\cM$ is an associative $\ast$-algebra. It contains, among others, the identity, the 
polynomials, the $\delta$ distribution and its derivatives. Then the relation
\begin{align}
  \lsb x^{\mu},x^{\nu}\rsb=&\imath\Theta^{\mu\nu}, 
\end{align}
often given as a definition of the Moyal space, holds in $\cM$ (but not in $\cA_{\Theta}$).

\subsubsection{ The \texorpdfstring{$\phi^{4}$}{phi4}-theory on \texorpdfstring{${\mathbb R}^{4}_{\theta}
$}{Moyal space}}

The simplest \encv{} model one may consider is the $\phi^{4}$-theory on the four-dimensional Moyal 
space. Its Lagrangian is the usual (commutative) one where the pointwise product is replaced by the 
Moyal one:
\begin{align}
  S[\phi] =&\int d^4x \Big( -\frac{1}{2} \partial_\mu \phi
\star \partial^\mu \phi  + \frac{1}{2} m^2
\,\phi \star \phi
+ \frac{\lambda}{4} \phi \star \phi \star \phi \star
\phi\Big)(x).\label{eq:phi4naif}
\end{align}
Thanks to the formula \eqref{eq:moyal-def}, this action can be explicitly computed. The interaction part is 
given by the corollary \ref{cor:int-Moyal}:
\begin{align}
  \int dx\, \phi^{\star
    4}(x)=&\int\prod_{i=1}^{4}dx_{i}\,\phi(x_{i})\,\delta(x_{1}-x_{2}+x_{3}-x_{4})e^{\imath\varphi},
  \label{eq:interaction-phi4}\\
  \varphi=&\sum_{i<j=1}^{4}(-1)^{i+j+1}x_{i}\wed x_{j}.\nonumber
\end{align}
The main characteristic of the Moyal product is its non-locality. But its non-commutativity implies that the 
vertex of the model \eqref{eq:phi4naif} is only invariant under cyclic permutation of the fields. This 
restricted invariance incites to represent the associated Feynman graphs with ribbon graphs. One can 
then make a clear distinction between planar and non-planar graphs. This will be detailed in section \ref
{sec:multi-scale-analysisMatrix}.

Thanks to the delta function in \eqref{eq:interaction-phi4}, the oscillation may be written in different ways:
\begin{subequations}
  \begin{align}
    \delta(x_{1}-x_{2}+x_{3}-x_{4})e^{\imath\varphi}=&\delta(x_{1}-x_{2}+x_{3}-x_{4})e^{\imath
      x_{1}\wed x_{2}+\imath x_{3}\wed x_{4}}\\
    =&\delta(x_{1}-x_{2}+x_{3}-x_{4})e^{\imath
      x_{4}\wed x_{1}+\imath x_{2}\wed x_{3}}\\
    =&\delta(x_{1}-x_{2}+x_{3}-x_{4})\exp\imath(x_{1}-x_{2})\wed(x_{2}-x_{3}).\label{eq:oscill-trans}
  \end{align} 
\end{subequations}
The interaction is real and positive\footnote{Another way to prove it is from \eqref{eq:Moyal-involution}, $
\overline{\phi^{\star 4}}=\phi^{\star 4}$.}:
\begin{align}
  &\int\prod_{i=1}^{4}dx_{i}\phi(x_{i})\,\delta(x_{1}-x_{2}+x_{3}-x_{4})e^{\imath\varphi}\label{eq:int-
positive}\\
  =&\int dk\lbt\int dxdy\,\phi(x)\phi(y)e^{\imath k(x-y)+\imath x\wed y}\rbt^{\!\!2}\in\R_{+}.\notag
\end{align}
It is also translation invariant as shows equation \eqref{eq:oscill-trans}.

The property \ref{prop:trace} implies that the propagator is the usual one: $\hat{C}(p)=1/(p^{2}+m^{2})$.

\subsubsection{UV/IR mixing}
\label{sec:uvir-mixing}

The non-locality of the $\star$-product allows to understand the discovery of Minwalla, Van Raamsdonk 
and Seiberg \cite{MiRaSe}. They showed that not only the model \eqref{eq:phi4naif} isn't finite in the UV 
but also it exhibits a new type of divergences making it non-renormalisable. In the article \cite{Filk1996dm}, Filk computed the Feynman rules corresponding to \eqref{eq:phi4naif}. He showed that 
the planar amplitudes equal the commutative ones whereas the non-planar ones give rise to oscillations 
coupling the internal and external legs. A typical example is the the non-planar tadpole:
\begin{align} 
  \raisebox{-0.4\totalheight}{\includegraphics{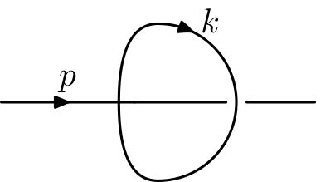}}&=\frac{\lambda}{12} \int \frac{d^4k}{(2\pi)^4} 
\frac{e^{ip_{\mu} k_{\nu}
      \Theta^{\mu\nu} }}{k^2 + m^2}\notag\\ 
  &=  \frac{\lambda}{48\pi^2}  \sqrt{\frac{m^2}{(\Theta p)^2}}  K_1(\sqrt{m^2
    (\Theta p)^2})\seq_{p\to 0}p^{-2}.
\end{align}
If $p\neq 0$, this amplitude is finite but, for small $p$, it diverges like $p^{-2}$. In other words, if we put 
an ultraviolet cut-off $\Lambda$ to the $k$-integral, the two limits $\Lambda\to\infty$ and $p\to 0$ do not 
commute. This is the UV/IR mixing phenomena. A chain of non-planar tadpoles, inserted in bigger 
graphs, makes divergent any function (with six points or more for example). But this divergence is not 
local and can't be absorbed in a mass redefinition. This is what makes the model non-renormalisable. 
We will see in sections \ref{sec:prop-et-renorm} and \ref{sec:direct-space} that the UV/IR mixing results 
in a coupling of the different scales of the theory. We will also note that we should distinguish different 
types of mixing.

The UV/IR mixing was studied by several groups. First, Chepelev and Roiban \cite{Chepelev2000hm} 
gave a power counting for different scalr models. They were able to identify the divergent graphs and to 
classify the divergences of the theories thanks to the topological data of the graphs. Then V.~Gayral \cite{Gayral2004cs} showed that UV/IR mixing is present on all isospectral deformations (they consist in 
curved generalisations of the Moyal space and of the \encv{} torus). For this, he considered a scalar 
model \eqref{eq:phi4naif} and discovered contributions to the effective action which diverge when the 
external momenta vanish. The UV/IR mixing is then a general characteristic of the \encv{} theories, at 
least on the deformations.
 
\subsection{The Grosse-Wulkenhaar breakthrough}

The situation remained so until H.~Grosse and R.~Wulkenhaar discovered a way to define a 
renormalisable \encv{} model. We will detail their result in section \ref{sec:multi-scale-analysisMatrix} but 
the main message is the following. By adding an harmonic term to the Lagrangian \eqref{eq:phi4naif},
\begin{align}
    S[\phi] =&\int d^4x \Big( -\frac{1}{2} \partial_\mu \phi
\star \partial^\mu \phi +\frac{\Omega^2}{2} (\tilde{x}_\mu \phi )\star (\tilde{x}^\mu \phi ) + \frac{1}{2} m^2
\,\phi \star \phi
+ \frac{\lambda}{4} \phi \star \phi \star \phi \star
\phi\Big)(x)\label{eq:phi4Omega}
\end{align}
where $\xt=2\Theta^{-1} x$ and the metric is Euclidean, the model, in four dimensions, is renormalisable 
at all orders of perturbation \cite{c}. We will see in section \ref{sec:direct-space} that this additional term 
give rise to an infrared cut-off and allows to decouple the different scales of the theory. The new model 
(\ref{eq:phi4Omega}), we call it $\Phi^{4}_{4}$, do not exhibit any mixing. This result is very important 
because it opens the way towards other \encv{} field theories. In the following, we will call \emph
{vulcanisation}\footnote{TECHNOL. Opération consistant à traiter le caoutchouc naturel ou 
synthétique par addition de soufre, pour en améliorer les propriétés mécaniques et la 
résistance aux variations de température, Trésor de la Langue Française informatisé, \href{http://
www.lexilogos.com/}{http://www.lexilogos.com/}.} the procedure consisting in adding a new term to a 
Lagrangian of a \encv{} theory in order to make it renormalisable.\\

The propagator $C$ of this $\Phi^{4}$ theory is the kernel of the inverse operator $-\Delta+\Omega^{2}
\xt^{2}+m^{2}$. It is known as the Mehler kernel \cite{simon79funct,toolbox05}
\begin{equation}
  \label{eq:Mehler}
  C(x,y)=\frac{\Omega^{2}}{\theta^{2}\pi^{2}}\int_{0}^{\infty}\frac{dt}{\sinh^{2}(2\Ot
  t)}\,e^{-\frac{\Ot}{2}\coth( 2\Ot t)(x-y)^{2}-\frac{\Ot}{2}\tanh(2\Ot t)(x+y)^{2}-m^{2}t}.
\end{equation}
Langmann and Szabo remarked that the quartic interaction with Moyal product is invariant under a 
duality transformation. It is a symmetry between momentum and direct space. The interaction part of the 
model \eqref{eq:phi4Omega} is (see equation \eqref{eq:int-Moyal-even})
\begin{align}
  S_{\text{int}}[\phi]=&\int d^{4}x\,\frac{\lambda}{4}(\phi\star\phi\star\phi\star\phi)(x)\\
  =&\int\prod_{a=1}^{4}d^{4}x_{a}\,\phi(x_{a})\,V(x_{1},x_{2},x_{3},x_{4})\label{eq:Vx}\\
  =&\int\prod_{a=1}^{4}\frac{d^{4}p_{a}}{(2\pi)^{4}}\,\hat{\phi}(p_{a})\,\hat{V}(p_{1},p_{2},p_{3},p_{4})\label
{eq:Vp}
  \intertext{with}
  V(x_{1},x_{2},x_{3},x_{4})=&\frac{\lambda}{4}\frac{1}{\pi^{4}\det\Theta}\delta(x_{1}-x_{2}+x_{3}-x_{4})
\cos(2(\Theta^{-1})_{\mu\nu}(x_{1}^{\mu}x_{2}^{\nu}+x_{3}^{\mu}x_{4}^{\nu}))\notag\\
  \hat{V}(p_{1},p_{2},p_{3},p_{4})=&\frac{\lambda}{4}(2\pi)^{4}\delta(p_{1}-p_{2}+p_{3}-p_{4})\cos(\frac 12
\Theta^{\mu\nu}(p_{1,\mu}p_{2,\nu}+p_{3,\mu}p_{4,\nu}))\notag
\end{align}
where we used a \emph{cyclic} Fourier transform: $\hat{\phi}(p_{a})=\int dx\,e^{(-1)^{a}\imath p_{a}x_{a}}
\phi(x_{a})$. The transformation
\begin{align}
  \hat{\phi}(p)\leftrightarrow\pi^{2}\sqrt{|\det\Theta|}\,\phi(x),&\qquad p_{\mu}\leftrightarrow\xt_{\mu}  
\end{align}
exchanges \eqref{eq:Vx} and \eqref{eq:Vp}. In addition, the free part of the model \eqref{eq:phi4naif} isn't 
covariant under this duality. The vulcanisation adds a term to the Lagrangian which restores the 
symmetry. The theory \eqref{eq:phi4Omega} is then covariant under the Langmann-Szabo duality:
\begin{align}
  S[\phi;m,\lambda,\Omega]\mapsto&\Omega^{2}\,S[\phi;\frac{m}{\Omega},\frac{\lambda}{\Omega^{2}},
\frac{1}{\Omega}].  
\end{align}
By symmetry, the parameter $\Omega$ is confined in $\lsb 0,1\rsb$. Let us note that for $\Omega=1$, 
the model is invariant.

\paragraph{}
\label{parag:interpHarm}
The interpretation of that harmonic term is not yet clear. But the vulcanisation procedure already allowed 
to prove the renormalisability of several other models on Moyal spaces such that $\phi^{4}_{2}$ \cite
{GrWu03-2}, $\phi^{3}_{2,4}$ \cite{Grosse2005ig,Grosse2006qv} and the LSZ models \cite
{Langmann2003if,Langmann2003cg,Langmann2002ai}. These last are of the type
\begin{align}
    S[\phi] =&\int d^nx \Big( \frac{1}{2} \bar{\phi}\star(-\partial_\mu+\xt_{\mu}+m)^{2}\phi
+ \frac{\lambda}{4} \bar{\phi} \star \phi \star \bar{\phi} \star\phi\Big)(x).\label{eq:LSZintro}
\end{align}
By comparison with \eqref{eq:phi4Omega}, one notes that here the additional term is formally equivalent 
to a fixed magnetic background. Deep is the temptation to interpret it as such. This model is invariant 
under the above duality and is exactly soluble. Let us remark that the complex interaction in \eqref
{eq:LSZintro} makes the Langmann-Szabo duality more natural. It doesn't need a cyclic Fourier 
transform. The $\phi^{3}$ have been studied at $\Omega=1$ where they also exhibit a soluble structure.

\subsection{The non-commutative Gross-Neveu model}
\label{sec:non-comm-gross}

Apart from the $\Phi^{4}_{4}$, the modified Bosonic LSZ model \cite{xphi4-05}
and supersymmetric theories, we now know several renormalizable \encv{} field
theories. Nevertheless they either are super-renormalizable ($\Phi^{4}_{2}$
\cite{GrWu03-2}) or (and) studied at a special point in the parameter space
where they are solvable ($\Phi^{3}_{2},\Phi^{3}_{4}$
\cite{Grosse2005ig,Grosse2006qv}, the LSZ models
\cite{Langmann2003if,Langmann2003cg,Langmann2002ai}). Although only
logarithmically divergent for parity reasons, the \encv{} Gross-Neveu model is a
just renormalizable quantum field theory as $\Phi^{4}_{4}$. One of its main
interesting features is that it can be interpreted as a non-local
Fermionic field theory in a constant magnetic background. Then apart from
strengthening the ``vulcanization'' procedure to get renormalizable \encv{} field
theories, the Gross-Neveu model may also be useful for the study of the
quantum Hall effect. It is also a good first candidate for a constructive
study \cite{Riv1} of a \encv{} field theory as Fermionic models are usually
easier to construct. Moreover its commutative counterpart being asymptotically
free and exhibiting dynamical mass generation
\cite{Mitter1974cy,Gross1974jv,KMR}, a study of the physics of this model
would be interesting.\\

The \encv{} Gross-Neveu model ($\GN$) is a Fermionic quartically interacting quantum field theory on 
the Moyal plane $\R^{2}_{\theta}$. The skew-symmetric matrix $\Theta$ is
\begin{align}
  \Theta=&
  \begin{pmatrix}
    0&-\theta\\\theta&0
  \end{pmatrix}.
\end{align}
The action is
\begin{align}\label{eq:actfunctGN}
  S[\psib,\psi]=&\int
  dx\lbt\psib\lbt-\imath\slashed{\partial}+\Omega\xts+m+\mu\,\gamma_{5}\rbt\psi+V_{\text{o}}(\psib,\psi)
  +V_{\text{no}}(\psib,\psi)\rbt(x)
\end{align}
where $\xt=2\Theta^{-1}x$, $\gamma_{5}=\imath\gamma^{0}\gamma^{1}$ and $V=V_{\text{o}}+V_{\text
{no}}$ is the interaction part given hereafter. The $\mu$-term appears at two-loop order. We use a 
Euclidean metric and the Feynman convention $\slashed{a}=\gamma^{\mu}a_{\mu}$. The $\gamma^{0}
$ and $\gamma^{1}$ matrices form a two-dimensional representation of the Clifford algebra $\{\gamma^
{\mu},\gamma^{\nu}\}=-2\delta^{\mu\nu}$. Let us remark that the $\gamma^{\mu}$'s are then skew-
Hermitian: $\gamma^{\mu\dagger}=-\gamma^{\mu}$.

\paragraph{Propagator}
The propagator corresponding to the action \eqref{eq:actfunctGN} is given by the following lemma:
\begin{lemma}[Propagator \cite{toolbox05}]\label{xpropa1GN}
  The propagator of the Gross-Neveu model is
  \begin{align}
    C(x,y)=&\int d\mu_{C}(\psib,\psi)\,\psi(x)\psib(y)=\lbt-\imath\slashed{\partial}+\Omega\xts+m\rbt^{-1}
(x,y)\\
    =&\ \int_{0}^{\infty}dt\, C(t;x,y),\notag\\
    C(t;x,y)=&\ -\frac{\Omega}{\theta\pi}\frac{e^{-tm^{2}}}{\sinh(2\Ot t)}\,
    e^{-\frac{\Ot}{2}\coth(2\Ot t)(x-y)^{2}+\imath\Omega x\wed y}\\
    &\times\lb\imath\Ot\coth(2\Ot t)(\xs-\ys)+\Omega(\xts-\yts)-m\rb
    e^{-2\imath\Omega t\gamma\Theta^{-1}\gamma}\notag
  \end{align}
  with $\Ot=\frac{2\Omega}{\theta}$ et $x\wed y=2x\Theta^{-1}y$.\\
We also have $e^{-2\imath\Omega t\gamma\Theta^{-1}\gamma}=\cosh(2\Ot t)\mathds{1}_{2}-\imath\frac
{\theta}{2}\sinh(2\Ot
  t)\gamma\Theta^{-1}\gamma$.
\end{lemma}
If we want to study a $N$-\emph{color} model, we can consider a propagator diagonal in these color 
indices.

\paragraph{Interactions}
Concerning the interaction part $V$, recall that (see corollary \ref{cor:int-Moyal}) 
for any $ f_{1}$, $f_{2}$, $f_{3}$, $f_{4}$ in $\cA_{\Theta}$,
\begin{align}
  \int dx\,\lbt f_{1}\star f_{2}\star f_{3}\star
  f_{4}\rbt(x)=&\frac{1}{\pi^{2}\det\Theta}\int\prod_{j=1}^{4}dx_{j}f_{j}(x_{j})\,
  \delta(x_{1}-x_{2}+x_{3}-x_{4})e^{-\imath\varphi},\label{eq:interaction-GN}\\
  \varphi=&\sum_{i<j=1}^{4}(-1)^{i+j+1}x_{i}\wed x_{j}.
\end{align}
This product is non-local and only invraiant under cyclic permutations of the fields. Then, contrary to the 
commutative Gross-Neveu model, for which there exits only one spinorial interaction, the $\GN$ model 
has, at least, six different interacitons: the \emph{orientable} ones
\begin{subequations}\label{eq:int-orient}
  \begin{align}
    V_{\text{o}}=\phantom{+}&\frac{\lambda_{1}}{4}\int
    dx\,\lbt\psib\star\psi\star\psib\star\psi\rbt(x)\label{eq:int-o-1}\\
    +&\frac{\lambda_{2}}{4}\int
    dx\,\lbt\psib\star\gamma^{\mu}\psi\star\psib\star\gamma_{\mu}\psi\rbt(x)\label{eq:int-o-2}\\
    +&\frac{\lambda_{3}}{4}\int
    dx\,\lbt\psib\star\gamma_{5}\psi\star\psib\star\gamma_{5}\psi\rbt(x),&\label{eq:int-o-3}
  \end{align}
\end{subequations}
where $\psi$'s and $\psib$'s alternate and the \emph{non-orientable} ones
\begin{subequations}\label{eq:int-nonorient}
  \begin{align}
    V_{\text{no}}=\phantom{+}&\frac{\lambda_{4}}{4}\int
    dx\,\lbt\psi\star\psib\star\psib\star\psi\rbt(x)\label{eq:int-no-1}&\\
    +&\frac{\lambda_{5}}{4}\int
    dx\,\lbt\psi\star\gamma^{\mu}\psib\star\psib\star\gamma_{\mu}\psi\rbt(x)\label{eq:int-no-2}\\
    +&\frac{\lambda_{6}}{4}\int
    dx\,\lbt\psi\star\gamma_{5}\psib\star\psib\star\gamma_{5}\psi\rbt(x).\label{eq:int-no-3}
  \end{align}
\end{subequations}
All these interactions have the same $x$ kernel thanks to the equation \eqref{eq:interaction-GN}. The 
reason for which we call these interactions orientable or not will be clear in section \ref{sec:direct-space}.

\section{Multi-scale analysis in the matrix basis}
\label{sec:multi-scale-analysisMatrix}

The matrix basis is a basis for Schwartz-class functions. In this basis, the Moyal product becomes a 
simple matrix product. Each field is then represented by an infinite matrix \cite{Gracia-Bondia1987kw,GrWu03-2,vignes-tourneret06:PhD}.

\subsection{A dynamical matrix model}
\label{sec:phi4-matrixbase}

\subsubsection{From the direct space to the matrix basis}
\label{sec:de-lespace-direct}

In the matrix basis, the action~\eqref{eq:phi4Omega} takes the form:
\begin{align}
  S[\phi]=&(2\pi)^{D/2}\sqrt{\det\Theta}\Big(\frac 12\phi\Delta\phi+\frac{\lambda}{4}\Tr\phi^{4}\Big)\label
{eq:SPhi4matrix}
\end{align}
where $\phi=\phi_{mn},\,m,n\in\N^{D/2}$ and
\begin{align}
    \Delta_{mn,kl}=&\sum_{i=1}^{D/2}\Big(\mu_{0}^{2}+\frac{2}{\theta}(m_{i}
    +n_{i}+1)\Big)\delta_{ml}\delta_{nk}\label{eq:formequadMatrixPhi4}\\
    &\hspace{-1.5cm}-\frac{2}{\theta}
    (1-\Omega^{2})\Big(\sqrt{(m_{i}+1)(n_{i}+1)}\,
    \delta_{m_{i}+1,l_{i}}\delta_{n_{i}+1,k_{i}}+\sqrt{m_{i}n_{i}}\,\delta_{m_{i}-1,l_{i}}
    \delta_{n_{i}-1,k_{i}}\Big)\prod_{j\neq i}\delta_{m_{j}l_{j}}\delta_{n_{j}k_{j}}.\notag
\end{align}
The (four-dimensional) matrix $\Delta$ represents the quadratic part of the Lagragian. The first difficulty 
to study the matrix model \eqref{eq:SPhi4matrix} is the computation of its propagator $G$ defined as the 
inverse of $\Delta$ :
\begin{align}
  \sum_{r,s\in\N^{D/2}}\Delta_{mn;rs}G_{sr;kl}
  =\sum_{r,s\in\N^{D/2}}G_{mn;rs}\Delta_{sr;kl}=\delta_{ml}\delta_{nk}.
\end{align}

Fortunately, the action is invariant under $SO(2)^{D/2}$ thanks to the form \eqref{eq:Thetamatrixbase} of 
the $\Theta$ matrix. It implies a conservation law
\begin{align}
  \Delta_{mn,kl}=&0\iff m+k\neq n+l.\label{eq:conservationindices}
\end{align}
The result is \cite{c,GrWu03-2}
\begin{align}
  \label{eq:propaPhimatrix}
  G_{m, m+h; l + h, l} 
&= \frac{\theta}{8\Omega} \int_0^1 d\alpha\,  
\dfrac{(1-\alpha)^{\frac{\mu_0^2 \theta}{8 \Omega}+(\frac{D}{4}-1)}}{  
(1 + C\alpha )^{\frac{D}{2}}} \prod_{s=1}^{\frac{D}{2}} 
G^{(\alpha)}_{m^s, m^s+h^s; l^s + h^s, l^s},
\\
 G^{(\alpha)}_{m, m+h; l + h, l}
&= \left(\frac{\sqrt{1-\alpha}}{1+C \alpha} 
\right)^{m+l+h} \sum_{u=\max(0,-h)}^{\min(m,l)}
   {\cal A}(m,l,h,u)\ 
\left( \frac{C \alpha (1+\Omega)}{\sqrt{1-\alpha}\,(1-\Omega)} 
\right)^{m+l-2u},\notag
\end{align}
where ${\cal A}(m,l,h,u)=\sqrt{\binom{m}{m-u}
\binom{m+h}{m-u}\binom{l}{l-u}\binom{l+h}{l-u}}$ and $C$ is a function in $\Omega$ : $C(\Omega)=
\frac{(1-\Omega)^2}{4\Omega}$. The main advantage of the matrix basis is that it simplifies the 
interaction part: $\phi^{\star 4}$ becomes $\Tr\phi^{4}$. But the propagator becomes very compllicated.\\

\begin{figure}[htb]
  \centering
  \includegraphics{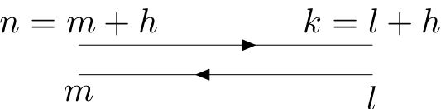}
  \caption{Matrix Propagator}
  \label{fig:propamatrix}
\end{figure}

Let us remark that the matrix model \eqref{eq:SPhi4matrix} is \emph{dynamical}: its quadratic part is not 
trivial. Usually, matrix models are \emph{local}.
\begin{defn}
A matrix model is called
\textbf{local} if $G_{mn;kl}=G(m,n)\delta_{ml}\delta_{nk}$ and \textbf{non-local} otherwise.
\end{defn}
In the matrix theories, the Feynman graphs are ribbon graphs. The propagator $G_{mn;kl}$ is then 
represented by the Figure \ref{fig:propamatrix}.
In a local matrix model, the propagator preserves the index values along the trajectories (simple lines).

\subsubsection{Topology of ribbon graphs}
\label{sec:topologie-des-graphes}

\begin{figure}[htbp]
  \centering 
  \subfloat[Planar]{{\label{fig:ribongraph1}}\includegraphics[scale=1]{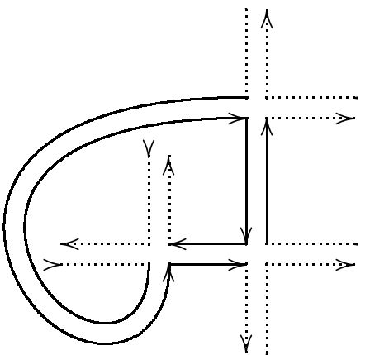}}\qquad
  \subfloat[Non-planar]{\label{fig:ribongraph2}\includegraphics[scale=1]{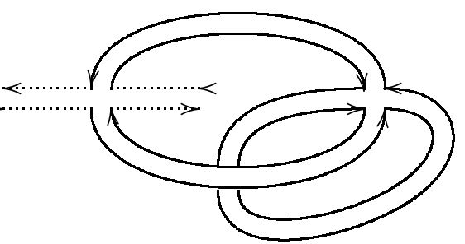}}
  \caption{Ribbon Graphs}
  \label{fig:ribbon-examples}
\end{figure}

\begin{figure}[hbtp]
  \centering
  \begin{minipage}[c]{3cm}
    \centering
    \includegraphics[width=3cm]{gt1.eps}
  \end{minipage}%
  \begin{minipage}[c]{2cm}
    \centering
    $ \Longrightarrow$
  \end{minipage}%
  \begin{minipage}[c]{2.6cm}
    \centering
    \includegraphics[width=2.6cm]{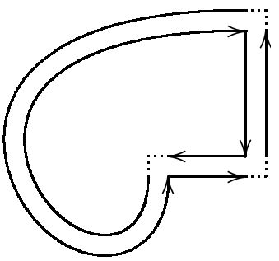}
  \end{minipage} \hspace{1cm}
   $\left .
     \begin{array}{c}
       $V=3$\\
       $I=3$\\
       $F=2$\\
       $B=2$
     \end{array}\rb
     \Longrightarrow\ g=0$\\
     \vspace{1cm}
     \begin{minipage}[c]{4cm}
       \centering
       \includegraphics[width=4cm]{gt3.eps}
     \end{minipage}%
     \begin{minipage}[c]{2cm}
       \centering
       $\Longrightarrow$
     \end{minipage}%
     \begin{minipage}[c]{3cm}
       \centering
       \includegraphics[width=3cm]{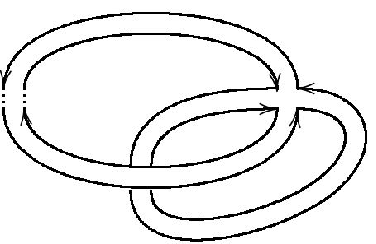}
     \end{minipage} \hspace{1cm}
     $\left .
       \begin{array}{c}
         $V=2$\\
         $I=3$\\
         $F=1$\\
         $B=1$
       \end{array}\rb
       \Longrightarrow\ g=1$
       \caption[Topological Data]{Topological Data of Ribbon Graphs}
       \label{fig:topo-ruban}
\end{figure}

The power counting of a matrix model depends on the topological data of its graphs. The figure \ref
{fig:ribbon-examples} gives two examples of ribbon graphs.
Each ribbon graph may be drawn on a two-dimensional manifold. Actually each graph defines the 
surface on which it is drawn. Let a graph $G$ with $V$ vertices, $I$ internal propagators (double lines) 
and $F$ faces (made of simple lines). The Euler characteristic
\begin{align}
  \chi=&2-2g=V-I+F\label{eq:Eulercar}
\end{align}
gives the genus $g$ of the manifold. One can make this clear by passing to the \textbf{dual graph}. The 
dual of a given graph $G$ is obtained by exchanging faces and vertices. The dual graphs of the $\Phi^
{4}$ theory are tesselations of the surfaces on which they are drawn. Moreover each direct face broken 
by exernal legs becomes, in the dual graph, a \textbf{puncture}. If among the $F$ faces of a graph, $B$ 
are broken, this graph may be drawn on a surface of genus $g=1-\frac 12(V-I+F)$ with $B$ punctures. 
The figure \ref{fig:topo-ruban} gives the topological data of the graphs of the figure \ref{fig:ribbon-examples}.

\subsection{Multi-scale analysis}
\label{sec:analyse-multi-echell-matrix}

In \cite{Rivasseau2005bh}, V.~R., R.~Wulkenhaar and F.~V.-T. used the multi-scale analysis to reprove 
the power counting of the \encv{} $\Phi^{4}$ theory.

\subsubsection{Bounds on the propagator}
\label{sec:bornes-sur-le}

Let $G$ a ribbon graph of the $\Phi^4_{4}$ theory with $N$ external legs, $V$ vertices, $I$ internal lines 
and $F$ faces. Its genus is then $g=1 - \frac 12(V-I+F)$. Four indices $\{m,n;k,l\} \in\N^2$ are associated 
to each internal line of the graph and two indices to each external line, that is to say $4I+2N =8V$ 
indices. But, at each vertex, the left index of a ribbon equals the right one of the neighbour ribbon. This 
gives rise to $4V$ independant identifications which allows to write each index in terms of a set $
\mathcal{I}$ made of $4V$ indices, four per vertex, for example the left index of each half-ribbon.\\

The graph amplitude is then
\begin{align}  
  A_{G} = \sum_{\mathcal{I}} \prod_{\delta \in G}
  G_{m_{\delta}(\mathcal{I}),n_{\delta}(\mathcal{I});k_{\delta}(\mathcal{I}),l_{\delta}(\mathcal{I})}\;
    \delta_{m_{\delta}-l_{\delta},n_{\delta}-k_{\delta}} \;,
\label{IG}
\end{align}
where the four indices of the propagator $G$ of the line $\delta$ are function of $\mathcal{I}$ and written
\\
$\{m_{\delta}(\mathcal{I}),n_{\delta}(\mathcal{I}); k_{\delta}(\mathcal{I}),l_{\delta}(\mathcal{I})\} $. We 
decompose each propagator, given by \eqref{eq:propaPhimatrix}:
\begin{align}  
G = \sum_{i=0}^{\infty}G^i\qquad \text{thanks to }\int_{0}^{1}d\alpha=\sum_{i=1}^{\infty}\int_{M^{-2i}}^{M^
{-2(i-1)}}d\alpha,\;M>1.
\end{align}
We have an associated decomposition for each amplitude
\begin{align}  
A_G &= \sum_{\mu} A_{G,\mu}\;,
\\
A_{G,\mu} &= \sum_{\mathcal{I}} \prod_{\delta \in G} G^{i_{\delta}}_{
m_{\delta}(\mathcal{I}),n_{\delta}(\mathcal{I});
k_{\delta}(\mathcal{I}),l_{\delta}(\mathcal{I})}  \;
\delta_{m_{\delta}(\mathcal{I})-l_{\delta}(\mathcal{I}),
n_{\delta}(\mathcal{I})-k_{\delta}(\mathcal{I})} \;,
\label{IGmu}
\end{align}
where $\mu=\{i_{\delta}\}$ runs over the all possible assignements of a positive integer $i_{\delta}$ to 
each line $\delta$. We proved the following four propositions:
\begin{prop}
\label{thm-th1}
For $M$ large enough, there exists a constant $K$ such that, for $\Omega\in [0.5,1]$, we have the 
uniform bound
\begin{equation} 
    \label{th1}
    G^i_{m,m+h;l+h,l}\les 
    KM^{-2i} e^{-\frac{\Omega}{3}M^{-2i}\|m+l+h\|}.
  \end{equation}
\end{prop}
\begin{prop}
\label{thm-th2}
For $M$ large enough, there exists two constants $K$ and $K_{1}$ such that, for $\Omega \in [0.5,1]$, 
we have the uniform bound
\begin{align} 
&   G^i_{m,m+h;l+h,l}
\nonumber
\\*
& \les K M^{-2i} e^{-\frac{\Omega}{4}M^{-2i}\|m+l+h\|} 
\prod_{s=1}^{\frac{D}{2}} \min\left( 1, 
\left(
    \frac{K_{1}\min(m^s,l^s,m^s+h^s,l^s+h^s)}{M^{2i}}
\right)^{\!\!\frac{|m^s-l^s|}{2}} \right).
\label{th2}
\end{align}
\end{prop}
This bound allows to prove that the only diverging graphs have either a constant index along the 
trajectories or a total jump of $2$.
\begin{prop}\label{prop:bound3}
For $M$ large enough, there exists a constant $K$ such that, for $\Omega \in [\frac 23,1]$, we have the 
uniform bound
\begin{align}
\sum_{l =-m}^p G^i_{m,p-l,p,m+l} &\les 
K M^{-2i} \,e^{-\frac{\Omega}{4} M^{-2i} (\|p\|+\|m\|) }\;.
\label{thsum}
\end{align}
\end{prop}
This bound shows that the propagator is almost local in the following sense: with $m$ fixed, the sum 
over $l$ doesn't cost anything (see Figure \ref{fig:propamatrix}). Nevertheless the sums we'll have to 
perform are entangled (a given index may enter different propagators) so that we need the following 
proposition.
\begin{prop}\label{prop:bound4}
For $M$ large enough, there exists a constant $K$ such that, for $\Omega \in [\frac 23,1]$, we have the 
uniform bound
  \begin{equation} \label{thsummax}
   \sum_{l=-m}^\infty\max_{p\ges\max(l,0)}G^i_{m,p-l;p,m+l}
\les  KM^{-2i}e^{-\frac{\Omega}{36}M^{-2i}\|m\|} \;.   
  \end{equation}
\end{prop}
We refer to \cite{Rivasseau2005bh} for the proofs of these four propositions.

\subsubsection{Power counting}
\label{sec:vari-indep}

About half of the $4V$ indices initially associated to a graph is determined by the external indices and 
the delta functions in (\ref{IG}). The other indices are summation indices. The power counting consists in 
finding which sums cost $M^{2i}$ and which cost $\mathcal{O}(1)$ thanks to (\ref{thsum}). The $M^{2i}$ 
factor comes from (\ref{th1}) after a summation over an index\footnote{Recall that each index is in fact 
made of two indices, one for each symplectic pair of $\R^{4}_{\theta}$.} $m\in\N^2$,
\begin{align}
\sum_{m^1,m^2=0}^\infty e^{- c M^{-2i}(m^1+m^2)} = \frac{1}{(1-e^{- c
    M^{-2i}})^2} = \frac{M^{4i}}{c^2} (1 + \mathcal{O}(M^{-2i})).
\label{summ1m2}
\end{align}

We first use the delta functions as much as possible to reduce the set $\mathcal{I}$ to a true minimal set 
$\cI'$ of independant indices. For this, it is convenient to use the dual graphs where the resolution of the 
delta functions is equivalent to a usual momentum routing.

The dual graph is made of the same propagators than the direct graph except the position of their 
indices. Whereas in the original graph we have $G_{mn;kl}=\raisebox{-1ex}{\includegraphics[scale=.6]
{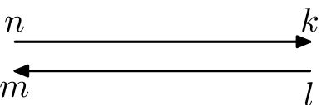}}$, the position of the indices in a dual propagator is
\begin{align}
G_{mn;kl} = \raisebox{-1ex}{\includegraphics[scale=.6]{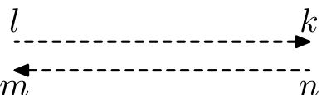}}\;.
\label{dual-assign}
\end{align}
The conservation $\delta_{l-m,-(n-k)}$ in (\ref{IG}) implies that the difference $l-m$ is conserved along 
the propagator. These differences behave like an \emph{angular momentum} and the conservation of 
the differences $\ell=l-m$ and $-\ell=n-k$ is nothing else than the conservation of the angular 
momentum thanks to the symmetry $SO(2)\times SO(2)$ of the action \eqref{eq:SPhi4matrix}:
\begin{align}
\raisebox{-1.5ex}{\includegraphics[scale=1]{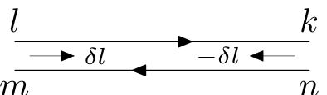}}&\qquad
l=m+\ell\;,~~ n=k+(-\ell).
\label{angmom}
\end{align}
The cyclicity of the vertices implies the vanishing of the sum of the angular momenta entering a vertex. 
Thus the angular momentum in the dual graph behaves exactly like the usual momentum in ordinary 
Feynman graphs.\\

We know that the number of independent momenta is exactly the number $L'$ ($=I-V'+1$ for a 
connected graph) of loops in the dual graph. Each index at a (dual) vertex is then given by a unique 
\emph{reference index} and a sum of momenta. If the dual vertex under consideration is an external one, 
we choose an external index for the reference index. The reference indices in the dual graph 
correspond to the loop indices in the direct graph. The number of summation indices is then $V'-B + L' = I 
+ (1-B)$ where $B\ges 0$ is the number of broken faces of the direct graph or the number of external 
vertices in the dual graph.\\

By using a well-chosen order on the lines, an optimized tree and a $L^{1}-L^{\infty}$ bound, one can 
prove that the summation over the angular momenta does not cost anything thanks to (\ref{thsum}). 
Recall that a connected component is a subgraph for which all internal lines have indices greater than 
all its external ones. The power counting is then:
\begin{align}
  A_G\les& K'^{V}\sum_{\mu}\prod_{i,k}M^{\omega(G^{i}_{k})}\\
  \text{with }\omega(G^{i}_{k})=&4(V'_{i,k}-B_{i,k})-2I_{i,k}=4(F_{i,k}-B_{i,k})-2I_{i,k}\\
  =&(4-N_{i,k})-4(2g_{i,k}+B_{i,k}-1)\notag
\end{align}
and $N_{i,k}$, $V_{i,k}$, $I_{i,k}=2V_{i,k}-\frac{N_{i,k}}{2}$, $F_{i,k}$ and $B_{i,k}$ are respectively the 
numbers of external legs, of vertices, of (internal) propagators, of faces and broken faces of the 
connected component $G^{i}_{k}$ ; $g_{i,k}= 1 - \frac{1}{2} (V_{i,k}-I_{i,k}+F_{i,k})$ is its genus. We have
\begin{thm}
\label{pc-slice}
The sum over the scales attributions $\mu$ converges if $\forall i,k,\, \omega(G^{i}_{k}) <0$.
\end{thm}
We recover the power counting obtained in \cite{GrWu03-1}.

From this point on, renormalisability of $\phi^4_4$ can proceed (however remark that it remains 
limited to $\Omega\in [0.5,1]$ by the technical estimates such as (\ref{th1}); this limitation
is overcome in the direct space method below).

The multiscale analysis allows to define the so-called effective expansion,
in between the bare and the renormalized expansion, which is optimal, 
both for physical and for constructive purposes \cite{Riv1}.
In this effective expansion only the subcontributions with all \textit{internal} scales higher 
than all \textit{external} scales have to be renormalised by counterterms 
of the form of the initial Lagrangian.

In fact only planar such subcontributions with a single external face
must be renormalised by such counterterms. This follows
simply from the the Grosse-Wulkenhaar moves defined in \cite{GrWu03-1}. These moves 
translate the external legs along the outer border of the planar graph, up to irrelevant corrections, 
until they all merge together into a term of the proper Moyal form, which is then absorbed in the effective 
constants definition. This requires only the estimates (\ref{th1})-(\ref{thsummax}), which were checked 
numerically in \cite{GrWu03-1}.

In this way the relevant and marginal counterterms can be shown to be of the Moyal type,
namely renormalise the parameters $\lambda$, $m$ and $\Omega$\footnote{The wave function 
renormalisation
i.e. renormalisation of the $\partial_\mu\phi\star \partial^\mu\phi$ term can be absorbed in a rescaling
of the field, called ``field strength renromalization.''}.

Notice that in the multiscale analysis there is no need for the relatively complicated use of 
Polchinski's equation \cite{Polch} made in \cite{GrWu03-1}. Polchinski's method, 
although undoubtedly very elegant for proving perturbative renormalisability
does not seem directly suited to constructive purposes, even in the case of simple 
Fermionic models such as the commutative Gross Neveu model, see e.g. \cite{DiRi}.

The BPHZ theorem itself for the renormalised expansion follows from finiteness
of the effective expansion by developing the
counterterms still hidden in the effective couplings. Its own finiteness can be checked 
e.g. through the standard classification of forests \cite{Riv1}. Let us however
recall once again that in our opinion the effective expansion, not the renormalised one
is the more fundamental object, both to describe the physics and to attack deeper mathematical 
problems, such as those of constructive theory \cite{Riv1,Riv2}.

The matrix base simplfies very much at $\Omega =1$, where the matrix propagator becomes diagonal,
i.e. conserves exactly indices. This property has been used for the general proof that the
beta function of the theory vanishes in the ultraviolet regime \cite{beta2-06}, leading to the
exciting perspective of a full non-perturbative construction of the model.

\subsection{Propagators on \encv{} space}
\label{sec:boite-outils}

We give here the results we get in \cite{toolbox05}. In this article, we computed the $x$-space and matrix 
basis kernels of operators which generalize the Mehler kernel \eqref{eq:Mehler}. Then we proceeded to 
a study of the scaling behaviours of these kernels in the matrix basis. This work is useful to study the 
\encv{} Gross-Neveu model in the matrix basis.

\subsubsection{Bosonic kernel}

The following lemma generalizes the Mehler kernel \cite{simon79funct}:
\begin{lemma}
  \label{HinXspace}Let $H$ the operator:
  \begin{equation}
    H=\frac{1}{2}\Big(-\Delta+ \Omega^2x^2-2\imath B(x_0\partial_1-x_1\partial_0)\Big).\label{eq:HinMat}
  \end{equation}
  The $x$-space kernel of $e^{-tH}$ is:
  \begin{equation}
    e^{-tH}(x,x')=\frac{\Omega}{2\pi\sinh\Omega t}e^{-A},\label{eq:propaxboson}
  \end{equation}
  \begin{equation}
    A=\frac{\Omega\cosh\Omega t}{2\sinh\Omega t}(x^2+x'^2)-
    \frac{\Omega\cosh Bt}{\sinh\Omega t}x\cdot x'-\imath
    \frac{\Omega\sinh Bt}{\sinh\Omega t}x\wedge x'.
  \end{equation}
\end{lemma}
\begin{rem}
  The Mehler kernel corresponds to $B=0$. The limit $\Omega=B\to 0$ gives the usual heat kernel.
\end{rem}
\begin{lemma}
  Let $H$ be given by \eqref{eq:HinMat} with $\Omega (B)\to 2\Omega/\theta (2B\theta)$. 
  Its inverse in the matrix basis is:
  \begin{align}
    H^{-1}_{m,m+h;l+h,l}=&\frac{\theta}{8\Omega} \int_0^1 d\alpha\,  
\dfrac{(1-\alpha)^{\frac{\mu_0^2 \theta}{8 \Omega}+(\frac{D}{4}-1)}}{  
(1 + C\alpha )^{\frac{D}{2}}} (1-\alpha)^{-\frac{4B}{8\Omega}h}\prod_{s=1}^{\frac{D}{2}} 
G^{(\alpha)}_{m^s, m^s+h^s; l^s + h^s, l^s},\label{eq:propbosonmatrix}
\\
 G^{(\alpha)}_{m, m+h; l + h, l}
&= \left(\frac{\sqrt{1-\alpha}}{1+C \alpha} 
\right)^{m+l+h} \sum_{u=\max(0,-h)}^{\min(m,l)}
   {\cal A}(m,l,h,u)\ 
\left( \frac{C \alpha (1+\Omega)}{\sqrt{1-\alpha}\,(1-\Omega)} 
\right)^{m+l-2u},\notag
\end{align}
where ${\cal A}(m,l,h,u)=\sqrt{\binom{m}{m-u}
\binom{m+h}{m-u}\binom{l}{l-u}\binom{l+h}{l-u}}$ and $C$ is a function of $\Omega$ : $C(\Omega)=
\frac{(1-\Omega)^2}{4\Omega}$.
\end{lemma}

\subsubsection{Fermionic kernel}

On the Moyal space, we modified the commutative Gross-Neveu model by adding a $\xts$ term (see 
lemma \ref{xpropa1GN}). We have
\begin{eqnarray}
G(x,y) &=& -\frac{\Omega}{\theta\pi}\int_{0}^{\infty}\frac{dt}{\sinh(2\Ot
t)}\, e^{-\frac{\Ot}{2}\coth(2\Ot t)(x-y)^{2}+\imath\Ot
x\wedge y}
\\ 
&&    \lb\imath\Ot\coth(2\Ot t)(\xs-\ys)+\Omega(\xts-\yts)- \mu \rb
e^{-2\imath\Ot
t\gamma^{0}\gamma^{1}}e^{-t\mu^{2}} \; . \nonumber
\end{eqnarray}
It will be useful to express $G$ in terms of commutators:
\begin{eqnarray}    
G(x,y)  &=&-\frac{\Omega}{\theta\pi}\int_{0}^{\infty}dt\,\lb \imath\Ot\coth(2\Ot
t)\lsb\xs, \Gamma^t  \rsb(x,y) \right.
\nonumber\\
&&
\left. +\Omega\lsb\xts, \Gamma^t \rsb(x,y)  -\mu \Gamma^t (x,y)  \rb
e^{-2\imath\Ot t\gamma^{0}\gamma^{1}}e^{-t\mu^{2}}, 
\label{xfullprop}
\end{eqnarray}
where
\begin{eqnarray}
\Gamma^t (x,y)  &=&
\frac{1}{\sinh(2\Ot t)}\,
e^{-\frac{\Ot}{2}\coth(2\Ot t)(x-y)^{2}+\imath\Ot x\wedge y}
\end{eqnarray}
with $\Ot=\frac{2\Omega}{\theta}$ and $x\wedge y=x^{0}y^{1}-x^{1}y^{0}$.\\

We now give the expression of the Fermionic kernel \eqref{xfullprop} in the matrix basis. The inverse of 
the quadratic form
\begin{equation}
\Delta=p^{2}+\mu^{2}+\frac{4\Omega^{2}}{\theta^2} x ^{2} +\frac{4B}{\theta}L_{2}
\end{equation}
is given by \eqref{eq:propbosonmatrix} in the preceeding section:
\begin{align}
  \Gamma_{m, m+h; l + h, l} 
  &= \frac{\theta}{8\Omega} \int_0^1 d\alpha\,  
  \dfrac{(1-\alpha)^{\frac{\mu^2 \theta}{8 \Omega}-\frac{1}{2}}}{  
    (1 + C\alpha )} 
  \Gamma^{\alpha}_{m, m+h; l + h, l}\,\label{eq:propinit}
\\
  \Gamma^{(\alpha)}_{m, m+h; l + h, l}
  &= \left(\frac{\sqrt{1-\alpha}}{1+C \alpha} 
  \right)^{m+l+h}\left( 1-\alpha\right)^{-\frac{Bh}{2\Omega}} \label{eq:propinit-b}\\
  &
  \sum_{u=0}^{\min(m,l)} {\cal A}(m,l,h,u)\ 
  \left( \frac{C \alpha (1+\Omega)}{\sqrt{1-\alpha}\,(1-\Omega)} 
  \right)^{m+l-2u}.\notag
\end{align}
The Fermionic propagator $G$ (\ref{xfullprop}) in the matrix basis may be deduced from the kernel \eqref
{eq:propinit}. We just set $B= \Omega$, add the missing term with $\gamma^0 \gamma^1$ and compute 
the action of $-\ps-\Omega\xts+\mu$ on $\Gamma$. We must then evaluate $\lsb x^{\nu},\Gamma\rsb$ in 
the matrix basis:
\begin{align}
  \lsb x^{0},\Gamma\rsb_{m,n;k,l}=&2\pi\theta\sqrt\frac{\theta}{8}\lb\sqrt{m+1}
  \Gamma_{m+1,n;k,l}-\sqrt{l}\Gamma_{m,n;k,l-1}+\sqrt{m}\Gamma_{m-1,n;k,l}
\right.\nonumber\\
&-\sqrt{l+1}\Gamma_{m,n;k,l+1}+\sqrt{n+1}\Gamma_{m,n+1;k,l}-\sqrt{k}
\Gamma_{m,n;k-1,l}\nonumber\\
&\left.+\sqrt{n}\Gamma_{m,n-1;k,l}-\sqrt{k+1}
  \Gamma_{m,n;k+1,l}\rb,\label{x0Gamma}\\
  \lsb
  x^{1},\Gamma\rsb_{m,n;k,l}=&2\imath\pi\theta\sqrt\frac{\theta}{8}\lb\sqrt{m+1}
  \Gamma_{m+1,n;k,l}-\sqrt{l}\Gamma_{m,n;k,l-1}-\sqrt{m}
  \Gamma_{m-1,n;k,l} \right.
\nonumber\\
&+\sqrt{l+1}\Gamma_{m,n;k,l+1}
-\sqrt{n+1}\Gamma_{m,n+1;k,l}+\sqrt{k}\Gamma_{m,n;k-1,l}
\nonumber\\
&\left.+\sqrt{n}\Gamma_{m,n-1;k,l}-\sqrt{k+1}\Gamma_{m,n;k+1,l}\rb.
  \label{x1Gamma}
\end{align}
This allows to prove:
\begin{lemma}
Let $G_{m,n;k,l}$ the kernel, in the matrix basis, of the operator\\
$\lbt\ps+\Omega\xts+\mu\rbt^{-1}$. We have:
\begin{align}
G_{m,n;k,l}=& 
-\frac{2\Omega}{\theta^{2}\pi^{2}} \int_{0}^{1} 
d\alpha\, G^{\alpha}_{m,n;k,l},\label{eq:propaFermiomatrix}
\\
G^{\alpha}_{m,n;k,l}=&\lbt\imath\Ot\frac{2-\alpha}{\alpha}\lsb\xs,
\Gamma^{\alpha}\rsb_{m,n;k,l}
+\Omega\lsb\slashed{\tilde{x}},\Gamma^{\alpha}\rsb_{m,n;k,l} - \mu\,\Gamma^{\alpha}_{m,n;k,l}\rbt
\nonumber\\
&\times\lbt\frac{2-\alpha}{2\sqrt{1-\alpha}}
\mathds{1}_{2}-\imath\frac{\alpha}{2\sqrt{1-\alpha}}\gamma^{0}\gamma^{1}
\rbt.\label{eq:matrixfullprop}
\end{align}
where $\Gamma^{\alpha}$ is given by (\ref{eq:propinit-b}) and the commutators bu the formulas (\ref
{x0Gamma}) and (\ref{x1Gamma}).
\end{lemma}
The first two terms in the equation (\ref{eq:matrixfullprop}) contain commutators and will be gathered 
under the name $G^{\alpha, {\rm comm}}_{m,n;k,l}$. The last term will be called $G^{\alpha, {\rm mass}}_
{m,n;k,l}$:
\begin{align}\label{commterm}
G^{\alpha, {\rm comm}}_{m,n;k,l}=& \lbt\imath\Ot\frac{2-\alpha}{\alpha}\lsb\xs,
\Gamma^{\alpha}\rsb_{m,n;k,l} +\Omega\lsb\slashed{\tilde{x}},\Gamma^{\alpha}\rsb_{m,n;k,l} \rbt  
\nonumber\\
&\times\lbt\frac{2-\alpha}{2\sqrt{1-\alpha}}
\mathds{1}_{2}-\imath\frac{\alpha}{2\sqrt{1-\alpha}}\gamma^{0}\gamma^{1} \rbt,\\
\notag\\
G^{\alpha, {\rm mass}}_{m,n;k,l}=& - \mu\, \Gamma^{\alpha}_{m,n;k,l}
\times\lbt\frac{2-\alpha}{2\sqrt{1-\alpha}}
\mathds{1}_{2}-\imath\frac{\alpha}{2\sqrt{1-\alpha}}\gamma^{0}\gamma^{1} \rbt.\label{massterm}
\end{align}

\subsubsection{Bounds}
\label{sec:bornes}

We use the multi-scale analysis to study the behaviour of the propagator \eqref{eq:matrixfullprop} and 
revisit more finely the bounds \eqref{th1} to \eqref{thsummax}. In a slice $i$, the propagator is
\begin{align}
  \Gamma^i_{m,m+h,l+h,l} 
  &=\frac{\theta}{8\Omega}  \int_{M^{-2i}}^{M^{-2(i-1)}} d\alpha\; 
  \dfrac{(1-\alpha)^{\frac{\mu_0^2 \theta}{8 \Omega}-\frac{1}{2}}}{  
    (1 + C\alpha )} 
  \Gamma^{(\alpha)}_{m, m+h; l + h, l}\;.
  \label{prop-slice-i}
\end{align}
\begin{eqnarray}
G_{m,n;k,l}&=& \sum_{i=1}^\infty G^i_{m,n;k,l} \ ; \ G^i_{m,n;k,l} = 
-\frac{2\Omega}{\theta^{2}\pi^{2}} \int_{M^{-2i}}^{M^{-2(i-1)}} 
d\alpha\, G^{\alpha}_{m,n;k,l}  \ .
\label{eq:matrixfullpropsliced}
\end{eqnarray}
Let $h= n-m$ and $p=l-m$. Without loss of generality, we assume $h \ges 0 $ and $p\ges 0$. Then the 
smallest index among $m,n,k,l$ is $m$ and the biggest is $k=m+h+p$. We have:
\begin{thm}\label{maintheorem}
Under the assumptions $h =n-m\ges 0$ and $p=l-m \ges 0$, there exists $K,c\in\R_{+}$ ($c$ depends 
on $\Omega$) such that the propagator of the \encv{} Gross-Neveu model in a slice $i$ obeys the bound
\begin{eqnarray}\label{mainbound1}  
\vert G^{i,{\rm comm}}_{m,n;k,l}\vert&\les&   
K M^{-i} \bigg( \chi(\alpha k>1)\frac{\exp \{- \frac{c p ^2  }{1+ kM^{-2i}}
- \frac{ c M^{-2i}}{1+k} (h - \frac{k}{1+C})^2 \}}{(1+\sqrt{ kM^{-2i}}) }  
\nonumber\\
&&+ \min(1,(\alpha k)^{p})e^{- c k M^{-2i} - c  p }\bigg).
\end{eqnarray}
The mass term is slightly different:
\begin{align} \label{mainbound2}  
\vert  G^{i,{\rm mass}}_{m,n;k,l}\vert\les&   
K M^{-2i} \bigg( \chi(\alpha k>1) \frac{\exp \{- \frac{c p ^2  }{1+ kM^{-2i}}
- \frac{ c M^{-2i}}{1+k} (h - \frac{k}{1+C})^2 \}}{1+\sqrt{ kM^{-2i}}}\notag
\\
&+\min(1,(\alpha k)^{p}) e^{- c k M^{-2i} - c  p }\bigg).
\end{align}
\end{thm}
\begin{rem}
  We can redo the same analysis for the $\Phi^{4}$ propagator and get
  \begin{equation}\label{eq:boundphi4}  
    G^i_{m,n;k,l}\les K M^{-2i}\min\lbt 1,(\alpha k)^{p}\rbt e^{-c(M^{-2i}k+p)}
  \end{equation}
which allows to recover the bounds \eqref{th1} to \eqref{thsummax}.
\end{rem}

\subsection{Propagators and renormalisability}
\label{sec:prop-et-renorm}

Let us consider the propagator \eqref{eq:propaFermiomatrix} of the \encv{} Gross-Neveu model. We saw 
in section \ref{sec:bornes} that there exists two regions in the space of indices where the propagator 
behaves very differently. In one of them it behaves as the $\Phi^{4}$ propagator and leads then to the 
same power counting. In the critical region, we have
\begin{align}
  G^{i}\les&K\frac{M^{-i}}{1+\sqrt{ kM^{-2i}}}\,e^{- \frac{c p ^2  }{1+ kM^{-2i}}
    -\frac{ c M^{-2i}}{1+k} (h - \frac{k}{1+C})^2}.
\end{align}
The point is that such a propagator does not allow to sum two reference indices with a unique line. This 
fact was useful in the proof of the power counting of the $\Phi^{4}$ model. This leads to a \emph
{renormalisable} UV/IR mixing.

\begin{figure}[!htbp]
  \begin{center}
    \subfloat[At scale $i$]{{\label{fig:sunseti}}\includegraphics[scale=.7]{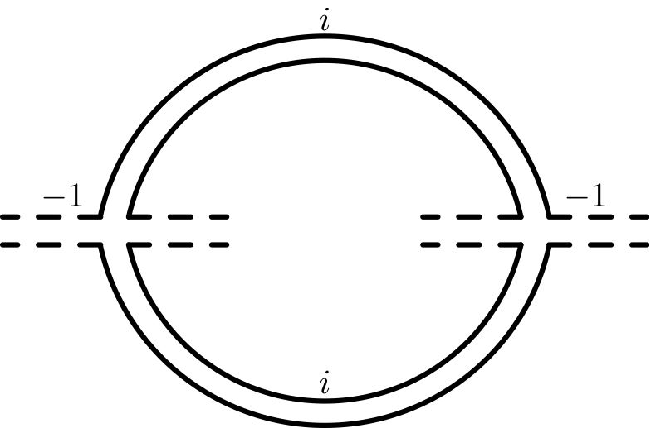}}\qquad
    \subfloat[At scale $j$]{{\label{fig:sunsetj}}\includegraphics[scale=.7]{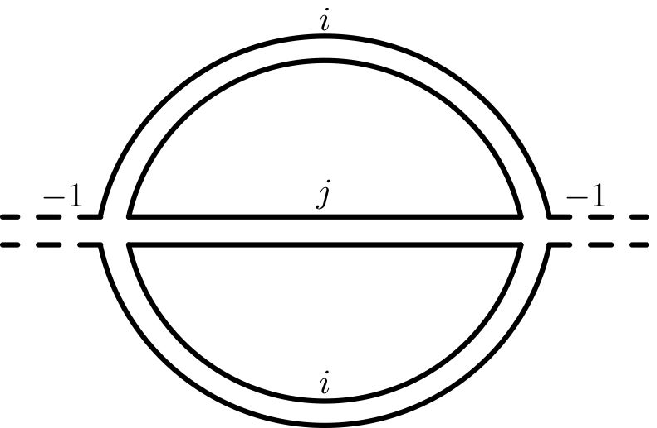}}
  \end{center}
  \caption{Sunset Graph}
  \label{figsunset}
\end{figure}

Let us consider the graph in figure \ref{fig:sunsetj} where the two external lines bear an index $i\gg 1$ 
and the internal one an index $j<i$. The propagator \eqref{eq:propaFermiomatrix} obeys the bound in 
Prop.~\eqref{thsum} which means that it is almost local. We only have to sum over one index per internal 
face.

On the graph of the figure \ref{fig:sunseti}, if the two lines inside are true external ones, the graph has 
two broken faces and there is no index to sum over. Then by using Prop.~\eqref{th1} we get $A_{G}\les 
M^{-2i}$. The sum over $i$ converges and we have the same behaviour as the $\Phi^{4}$ theory, that is 
to say the graphs with $B\ges 2$ broken faces are finite. But if these two lines belongs to a line of scale 
$j<i$ (see figure \ref{fig:sunsetj}), the result is different. Indeed, at scale $i$, we recover the graph of 
figure \ref{fig:sunseti}. To maintain the previous result ($M^{-2i}$), we should sum the two indices 
corresponding to the internal faces with the propagator of scale $j$. This is not possible. Instead we 
have:
\begin{align}
\sum_{k,h}M^{-2i-j}\,e^{-M^{-2i}k}\frac{e^{-\frac{ c M^{-2j}}{1+k} 
(h - \frac{k}{1+C})^2}}{1+\sqrt{ kM^{-2j}}}\les KM^{j}.
\end{align}
The sum over $i$ diverges logarithmically. The graph of figure \ref{fig:sunseti} converges if it is linked to 
true exernal legs et diverges if it is a subgraph of a graph at a lower scale. The power counting depends 
on the scales lower than the lowest scale of the graph. It can't then be factorized into the connected 
components: this is UV/IR mixing.\\

Let's remark that the graph of figure \ref{fig:sunseti} is not renormalisable by a counter-term in the 
Lagrangian. Its logarithmic divergence can't be absorbed in a redefinition of a coupling constant. 
Fortunately the renormalisation of the two-point graph of figure \ref{fig:sunsetj} makes the four-point 
subdivergence finite \cite{RenNCGN05}. This makes the \encv{} Gross-Neveu model renormalisable.

\section{Direct space}
\label{sec:direct-space}

We want now to explain how the power counting analysis
can be performed  in direct space, and the ``Moyality'' of the necessary counterterms
can be checked by a Taylor expansion which is a generalization of the one used in direct commutative 
space.

In the commutative case there is translation invariance, hence each propagator depends on a single
difference variable which is short in the ultraviolet regime; in the non-commutative case the propagator
depends both of the difference of end positions, which is again short in the uv regime, but also of the 
sum which is long in the uv regime, considering the explicit form (\ref{eq:Mehler}) of the Mehler kernel.

This distinction between short and long variables is at the basis of the  power counting analysis
in direct space.

\subsection{Short and long variables}

Let $G$ be an arbitrary connected graph. 
The amplitude associated with this graph is in direct space
(with hopefully self-explaining notations):
\begin{align}
A_G=&\int \prod_{v,i=1,...4} dx_{v,i} \prod_l dt_l     \\
& \prod_v \left[ \delta(x_{v,1}-x_{v,2}+x_{v,3}-x_{v,4})e^{\imath
\sum_{i<j}(-1)^{i+j+1}x_{v,i}\theta^{-1} x_{v,j}} \right] \prod_l C_l \; , 
\nonumber  \\
C_l=&  
\frac{\Omega^2}{[2\pi\sinh(\Omega t_l)]^2}e^{-\frac{\Omega}{2}\coth(\Omega 
t_l)(x_{v,i(l)}^{2}+x_{v',i'(l)}^{2})
+\frac{\Omega}{\sinh(\Omega t_l)}x_{v,i(l)} . x_{v',i'(l)}   - \mu_0^2 t_l}\; .\nonumber
\label{amplitude1}
\end{align} 

For each line $l$ of the graph joining positions $x_{v,i(l)}$ and $x_{v',i'(l)}$, 
we choose an orientation and we define 
the  ``short'' variable $u_l=x_{v,i(l)}-x_{v',i'(l)}$ and the 
``long'' variable $v_l=x_{v,i(l)}+x_{v',i'(l)}$.

With these notations, defining $\Omega t_l=\alpha_l$, the propagators in our graph can be 
written as:
\begin{equation}
\int_{0}^{\infty} \prod_l \frac{\Omega d\alpha_l}{[2\pi\sinh(\alpha_l)]^2}
e^{-\frac{\Omega}{4}\coth(\frac{\alpha_l}{2})
{ u_l^2}- \frac{\Omega}{4}\tanh(\frac{\alpha_l}{2})
{ v_l^2}  - \frac{\mu_0^2}{\Omega} \alpha_l}\; .\label{tanhyp}
\end{equation} 

As in matrix space we  can slice each propagator according to the size of its $\alpha$ parameter
and obtain the multiscale represenation of each Feynman amplitude:

\begin{align}
A_G=& \sum_{\mu}  A_{G,\mu}\quad, \quad A_{G,\mu} = \int \prod_{v,i=1,...4} dx_{v,i} \prod_l C_{l}^{i_
{\mu}(l)}(u_{l},v_{l})\label{amplitude2}\\ 
&\prod_v \left[ \delta(x_{v,1}-x_{v,2}+x_{v,3}-x_{v,4})e^{\imath
\sum_{i<j}(-1)^{i+j+1}x_{v,i}\theta^{-1} x_{v,j}} \right]\nonumber\\
 C^i (u,v) =& \int_{M^{-2i}}^{M^{-2(i-1)}} 
\frac{\Omega d\alpha}{[2\pi\sinh(\alpha)]^2}
e^{-\frac{\Omega}{4}\coth(\frac{\alpha}{2})
{ u^2}- \frac{\Omega}{4}\tanh(\frac{\alpha}{2})
{ v^2}  - \frac{\mu_0^2}{\Omega} \alpha}\; ,
\end{align} 
where $\mu$ runs over scales attributions $\{i_{\mu}(l) \}$ for each line $l$ of the graph, 
and the sliced propagator $C^i$ in slice $i \in {\mathbb N}$
obeys the crude bound:
\begin{lemma} For some constants $K$ (large) and $c$ (small):
\begin{equation}\label{eq:propbound-phi4}
C^i (u,v) \les K M^{2i}e^{-c [ M^{i}\Vert u \Vert + M^{-i}\Vert v\Vert ] }
\end{equation} 
(which a posteriori justifies the terminology of ``long'' and ```short'' variables).
\end{lemma}

The proof is elementary.

\subsection{Routing, Filk moves}
\label{sec:routing-filk-moves}
\subsubsection{Oriented graphs}

\begin{figure}[!htb]
  \centering
  \subfloat[Orientation of a tree]{\label{otree}\includegraphics[scale=0.7]{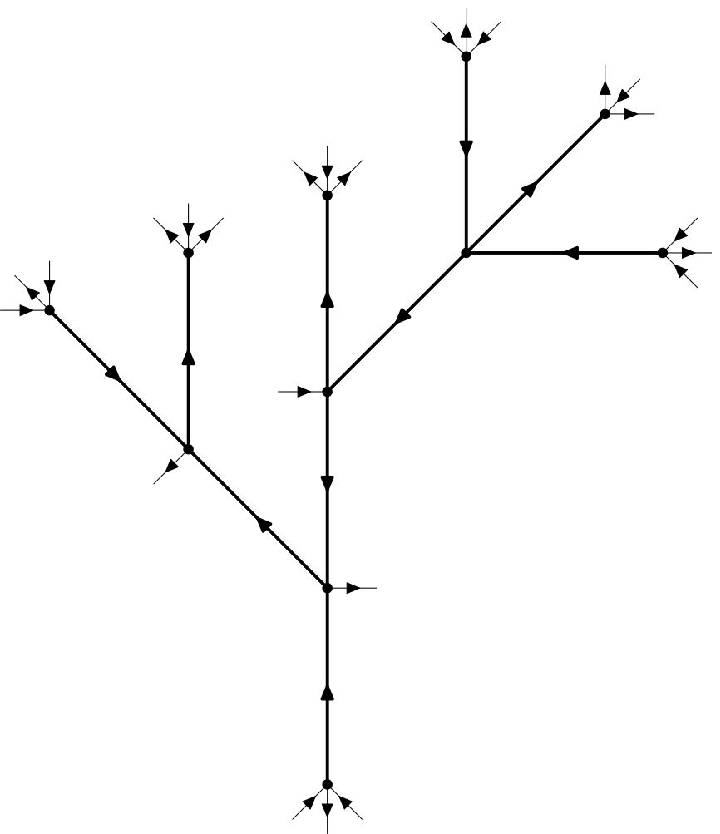}}\qquad
  \subfloat[A non-orientable graph]{\label{nono}\includegraphics[scale=0.6]{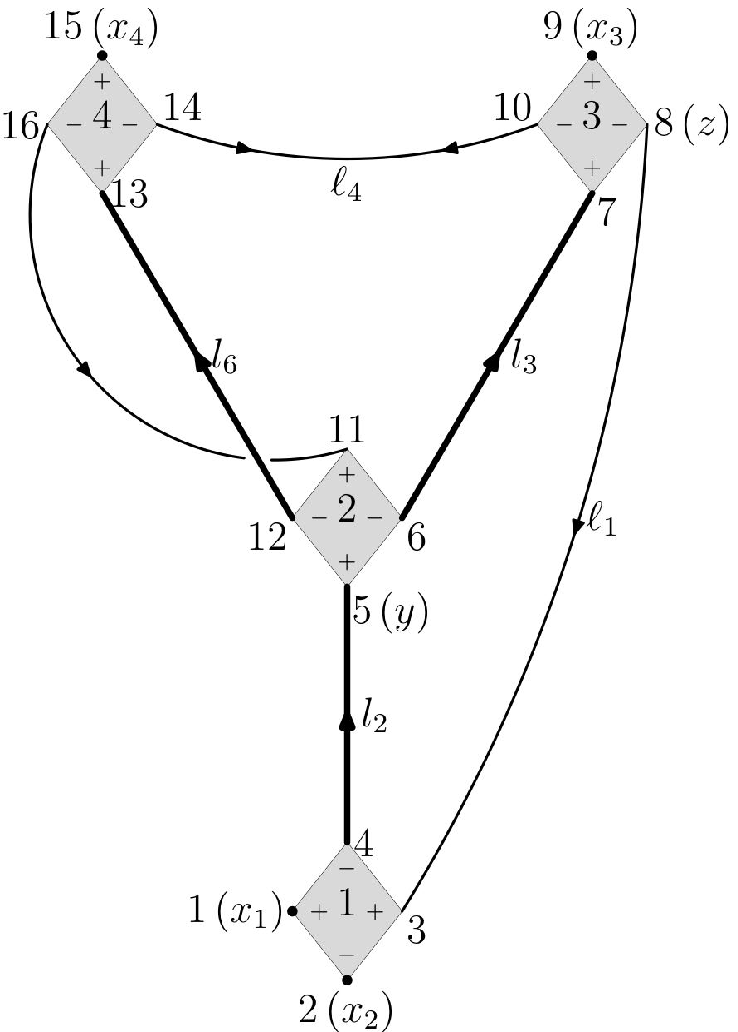}}
  \caption{Orientation}
  \label{fig:orientation}
\end{figure}

We pick a tree $T$ of lines of the graph, hence connecting all vertices, pick with a root vertex and build 
an \textit{orientation} of all the lines of the graph in an inductive way. Starting from an arbitrary 
orientation of a field at the 
root of the tree, we climb in the tree and at each vertex of the tree 
we impose cyclic order to alternate entering and exiting tree lines and loop half-lines, as in figure \ref
{otree}.
Then we look at the loop lines. If every loop lines consist in the contraction of an 
entering and an exiting line, the graph is called orientable. Otherwise we call it non-orientable
as in figure \ref{nono}.

\subsubsection{Position routing}

There are $n$ $\delta$ functions in an amplitude with $n$ vertices,
hence $n$ linear equations for the $4n$ positions,
one for each vertex. The {\it position routing} associated to the tree $T$ 
solves this system by passing to another equivalent system of $n$ linear equations,
one for each branch of the tree. This is a triangular change of variables,
of Jacobian 1. This equivalent system is obtained by summing the arguments of the 
$\delta$ functions of the vertices in each branch. This change of variables is exactly the 
$x$-space analog of the resolution of momentum conservation called \textit{momentum routing}
in the standard physics literature of commutative field theory, except that 
one should now take care of the additional $\pm$ cyclic signs.

One can prove \cite{xphi4-05} that the rank of the system of $\delta$ functions 
in an amplitude with $n$ vertices is
\begin{itemize}
\item $n-1$ if the graph is orientable
\item $n$ if the graph is non-orientable
\end{itemize}

The position routing change of variables is summarized by the following lemma:

\begin{lemma}[Position Routing]
We have, calling $I_G$ the remaining integrand in (\ref{amplitude2}):
\begin{align}
A_G =& \int \Big[ \prod_v  \big[ \delta(x_{v,1}-x_{v,2}+x_{v,3}-x_{v,4})\big] \, \Big]\;
I_G(\{x_{v,i}  \}  )   \\
=& \int \prod_{b}
\delta \left(   \sum_{l\in T_b \cup L_b }u_{l} + \sum_{l\in L_{b,+}}v_{l}-\sum_{l\in L_{b,-}}v_{l}
+\sum_{f\in X_b}\epsilon(f) x_f \right) I_G(\{x_{v,i}  \}), \nonumber 
\end{align} 
where $\epsilon(f)$ is $\pm 1$ depending on whether the field $f$ enters or exits the branch.
\end{lemma}

We can now use the system of delta functions to eliminate variables. It is of course better to eliminate
long variables as their integration costs a factor $M^{4i}$ whereas the integration of a short
variable brings $M^{-4i}$. Rough power counting, neglecting all oscillations of the vertices
leads therefore, in the case of an orientable graph with $N$ external fields, $n$ internal vertices
and $l= 2n - N/2$ internal lines at scale $i$ to:
\begin{itemize}
\item a factor $M^{2i(2n - N/2)}$ coming from the $M^{2i}$ factors for each line of scale $i$
  in (\ref{eq:propbound-phi4}),
\item a factor $M^{-4i(2n - N/2)}$ for  the $l = 2n - N/2$ short variables integrations,
\item a factor $M^{4i (n - N/2 +1)}$ for the long variables after eliminating $n-1$ of them 
using the delta functions.
\end{itemize}
The total factor is therefore $M^{-(N-4)i}$, the ordinary scaling of $\phi^4_4$, which means that
only two and four point subgraphs ($N \les 4)$ diverge when $i$ has to be summed.

In the non-orientable case, we can eliminate one additional long variable since the rank of
the system of delta functions is larger by one unit! Therefore we get a power counting bound
$M^{-Ni}$, which proves that  only {\it orientable} graphs may diverge.

In fact we of course know that not all {\it orientable}  two and four point subgraphs diverge
but only the planar ones with a single external face. 
(It is easy to check that all such planar graphs are indeed orientable).

Since only these planar subgraphs with a single external face can be renormalised by Moyal 
counterterms, we need to prove that orientable, non-planar graphs or orientable planar graphs with 
several external faces have in fact a better power  counting than this crude estimate. 
This can be done only by exploiting their vertices oscillations. 
We explain now how to do this with minimal effort.

\subsubsection{Filk moves and rosettes}

Following Filk \cite{Filk1996dm}, we can contract all lines of a spanning tree $T$
and reduce $G$ to a single vertex with ``tadpole loops'' called a ``rosette graph''.
This rosette is a cycle (which is the border of the former tree) bearing loops lines on it (see figure \ref
{fig:rosette}):
\begin{figure}[!htb]
  \centering
    \includegraphics[scale=0.4]{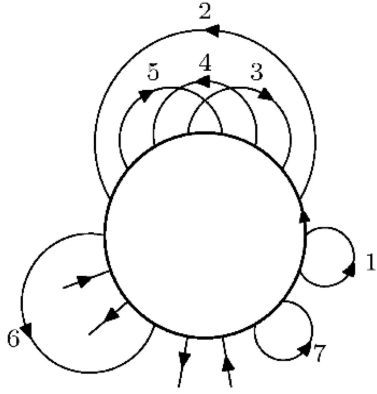}
    \caption{A rosette}
    \label{fig:rosette}
\end{figure}
Remark that the rosette can also be considered 
as a big vertex, with $r=2n+2$ fields, on which $N$
are external fields with external variables $x$ and $2n+2-N$ are loop fields for the corresponding
$n+1-N/2$ loops. When the graph is orientable, the rosette is also orientable,
which means that turning around the rosette the lines alternatively enter and exit.
These lines correspond to the contraction of the fields on the border of the tree $T$
before the Filk contraction, also called the ``first Filk move''.

\subsubsection{Rosette factor}

We start from the root and turn around the tree in the trigonometrical sense. We number
separately all the fields as $1,\dots,2n+2$ and all 
the tree lines as $1,\dots,n-1$ in the order they are met.
\begin{lemma}
The rosette contribution after a complete first Filk reduction is exactly:
\begin{equation}
\delta(v_1-v_2+\dots-v_{2n+2}+\sum_{l\in T}u_l)
e^{i VQV + iURU + iUSV}
\end{equation}
where the $v$ variables are the long or external variables
of the rosette, counted with their signs,
and the quadratic oscillations for these variables is
\begin{equation}
VQV= \sum_{0\les i<j\les r}(-1)^{i+j+1}v_i\theta^{-1} v_j 
\end{equation}
\end{lemma}

We have now to analyze in detail this quadratic oscillation of the remaining long loop variables 
since it is essential to improve power counting. We can neglect the
secondary oscillations $URU$ and $USV$ which imply short variables.

The second Filk reduction \cite{Filk1996dm} further simplifies the rosette factor by erasing the
loops of the rosette which do not cross any other loops or arch over external fields. 
It can be shown that the loops which disappear in this operation correspond to those 
long variables who do not appear in the quadratic form $Q$.

Using the remaining {\it oscillating factors}
one can prove that non-planar graphs with genus larger than one or with more than
one external face {\it do not diverge}.

The basic mechanism to improve the
power counting of a single non-planar subgraph is the following:
\begin{align}\label{gainoscill}
&\int dw_1dw_2 e^{-M^{-2i_1}w_1^2-M^{-2i_2}w_2^2
- iw_1\theta^{-1}w_2+w_1 . E_1(x,u)+w_2 E_2(x,u)}
\nonumber\\
=& \int dw'_1dw'_2 e^{-M^{-2i_1}(w_1')^2
-M^{-2i_2}(w'_2)^2 +iw'_1\theta^{-1}w'_2 + (u,x)Q(u,x)}
\nonumber\\
=&  K  M^{4i_1} \int dw'_2
e^{- (M^{2i_1}+ M^{-2i_2})(w'_2)^2 }=
K M^{4i_1}M^{-4i_2} \; .
\end{align} 
In these equations we used for simplicity $M^{-2i}$ 
instead of the correct but more complicated factor $(\Omega /4) \tanh (\alpha /2 )$
(of course this does not change the argument) and we performed
a unitary linear change of variables $w'_1 = w_1 + \ell_1 (x, u)$, $w'_2 = w_2 + \ell_2 (x, u)$
to compute the oscillating $w'_1$ integral. The gain in (\ref{gainoscill}) is
$M^{-8i_2}$, which is the difference between $M^{-4i_2}$ and 
the normal factor $M^{4i_2}$ that the $w_2$ integral would have cost if
we had done it with the regular $e^{-M^{-2i_2}w_2^2}$ factor for long variables. 
To maximize this gain we can assume $i_1 \les i_2$.

This basic argument must then be generalized to each non-planar 
subgraph in the multiscale analysis, which is possible.

Finally it remains to consider the case of subgraphs which are planar orientable
but with more than one external face. In that case there are no crossing loops in the rosette but
there must be at least one loop line arching over a non trivial subset
of external legs (see e.g. line $6$ in figure \ref{fig:rosette}). We have then a non trivial integration over at 
least one external variable, called $x$, of at least one long loop variable called $w$. This ``external'' $x$ 
variable without the oscillation improvement 
would be integrated with a test function of scale 1 (if it is a true external line of scale $1$)
or better (if it is a higher long loop variable)\footnote{Since the loop line arches 
over a non trivial (i.e. neither full nor empty) subset
of external legs of the rosette, the variable $x$ cannot be the full combination 
of external variables in the ``root'' $\delta$ function.}. But we get now
\begin{align}\label{gainoscillb}
&\int dx dw e^{-M^{-2i}w^2
- iw\theta^{-1}x  +w.E_1(x',u)}
\nonumber\\
=&  K  M^{4i} \int dx 
e^{-M^{+2i} x^2 }=
K' \ ,
\end{align} 
so that a factor $M^{4i}$ in the former bound becomes $\cO(1)$ hence is improved by $M^{-4i}$.

In this way we can reduce the convergence of the multiscale analysis to the
problem of renormalisation of planar two- and four-point subgraphs 
with a single external face, which we treat in the next section.

Remark that the power counting obtained in this way is still not optimal. To get the same level 
of precision than with the matrix base requires e.g. to display $g$ independent improvements
of the type (\ref{gainoscill}) for a graph of genus $g$. This is doable but basically requires
a reduction of the quadratic form $Q$ for single-faced rosette (also called ``hyperrosette'')
into $g$ standard symplectic blocks
through the so-called ``third Filk move'' introduced in \cite{gurauhypersyman}.
We return to this question in section \ref{hyperbo}.

\subsection{Renormalisation}

\subsubsection{Four-point function}

Consider the amplitude of a four-point graph $G$ which in the multiscale expansion
has all its internal scales higher than its four external scales.

The idea is that one should compare its amplitude to a similar amplitude with a
``Moyal factor'' $\exp\Big(
2\imath\theta^{-1}\lbt x_{1}\wedge x_{2}+x_{3}\wedge x_{4}\rbt\Big)\delta(\Delta)$
factorized in front, where $\Delta= x_1- x_{2}+x_{3} - x_{4}$.
But precisely because the graph is planar with a single external face
we understand that the external positions $x$ only couple to \textit{short variables} $U$ 
of the internal amplitudes through the 
global delta function and the oscillations. Hence we can break this coupling 
by a systematic Taylor expansion to first order. This separates a piece proportional to 
``Moyal factor'', then absorbed into the effective coupling constant, and a remainder 
which has at least one additional small factor which gives him improved power counting.

This is done by expressing the amplitude for a graph with $N=4$, $g = 0$ and $B = 1$ as:
\begin{align}
A(G)(x_1, x_2, x_3, x_4) =&\int  {\exp\Big(
2\imath\theta^{-1}\lbt x_{1}\wedge x_{2}+x_{3}\wedge x_{4}\rbt\Big)}
\prod_{\ell\in T^{i}_{k} }  du_{\ell} \ C_{\ell}(u_\ell, U_\ell, V_\ell) 
\nonumber\\
&
\bigg[ \prod_{l \in G^{i}_{k} \, \ l \not \in T}  du_{l}  d v_{l} C_{l}(u_l, v_l) \bigg]
\ e^{\imath URU+\imath USV}
\\
&\hspace{-2cm}
\Bigg\{ {\delta(\Delta)}
+ \int_{0}^{1}dt\bigg[ \mathfrak{U}\cdot \nabla \delta(\Delta+t\mathfrak{U})
+\delta(\Delta+t\mathfrak{U}) [\imath XQU  + {\mathfrak R}' (t)]  \bigg] 
e^{\imath tXQU + {\mathfrak R}(t)}  \Bigg\} \ .\nonumber
\end{align}
where
$C_{\ell}(u_\ell, U_\ell, V_\ell) $ is the propagator taken at $X_\ell=0$, $\mathfrak{U}= \sum_\ell u_\ell$
and ${\mathfrak R}(t)$ is a correcting term involving $\tanh \alpha_\ell [X.X + X.(U+V)]$.

The {first term} is of the initial $\int Tr \phi \star \phi \star \phi \star \phi $ form.
The rest no longer diverges, since the $U$ and ${\mathfrak R}$
provide the necessary small factors.

\subsubsection{Two-point function}

Following the same strategy we have to Taylor-expand the coupling
between external variables and $U$ factors in two point planar graphs with a single external face 
to \textit{third order} and some non-trivial symmetrization of the terms
acording to the two external arguments to cancel some odd contributions.
The corresponding factorized relevant and marginal contributions 
can be then shown to give rise only to

\begin{itemize}
\item A mass counterterm,
\item A wave function counterterm,
\item An {harmonic potential counterterm}.
\end{itemize}
and the remainder has convergent power counting.
This concludes the construction of the effective expansion in this direct space multiscale analysis.

Again the BPHZ theorem itself for the renormalised expansion follows by developing the
counterterms still hidden in the effective couplings and its finiteness
follows from the standard classification of forests. 
See however the remarks 
at the end of section \ref{sec:vari-indep}.

Since the bound (\ref{eq:propbound-phi4}) works for any $\Omega \ne 0$,
an additional bonus of the $x$-space
method is that it proves renormalisability of the model for any $\Omega$ in $]0,1]$\footnote{The case $
\Omega$ in $[1,+\infty[$ is irrelevant since it can be rewritten by LS duality as an equivalent
model with $\Omega$ in $]0,1]$.},
whether the matrix method proved it only for $\Omega$ in $]0.5,1]$.

\subsubsection{The Langmann-Szabo-Zarembo model}  

It is a four-dimensional theory of a Bosonic complex field defined by the action
\begin{align}
S=&\int \frac{1}{2} \bar \phi (- D^\mu D_\mu + \Omega^2 x^2  )\phi + \lambda 
\bar \phi \star \phi\star\bar \phi \star \phi
\end{align}
where $D^\mu= \imath\partial_\mu+B_{\mu \nu}x^\nu$ is the covariant derivative in a magnetic field $B$.

The interaction $\bar \phi \star \phi\star\bar \phi \star \phi$
ensures that perturbation theory contains only orientable graphs.
For $\Omega >0$ the $x$-space propagator still decays as in the ordinary $\phi^4_4$ case 
and the model has been shown renormalisable by an easy extension of the methods of the previous 
section \cite{xphi4-05}.

However at $\Omega =0$, there is no longer any harmonic potential in addition 
to the covariant derivatives and the bounds are lost. 
Models in this category are called ``critical''.

\subsubsection{Critical models}  

Consider the $x$-kernel of the operator
\begin{align}
H^{-1}&=\lbt p^2 + \Omega^2\xt^2 - 2\imath B\lbt x^0p_1-x^1p_0 \rbt \rbt^{-1}\\
H^{-1}(x,y)&=\frac{\Ot}{8\pi}\int_0^\infty \frac{dt}{\sinh(2\Ot t)}\, \exp\lbt -\frac{\Ot}{2}\frac{\cosh(2Bt)}{\sinh(2
\Ot t)} (x-y)^2\right.\\
&\hspace{1.5cm}\left. -\frac{\Ot}{2}{\frac{\cosh(2\Ot t)-\cosh(2Bt)}{\sinh(2\Ot t)}(x^2+y^2)} \right.\\
&\hspace{1.5cm}\left. +2\imath\Ot{\frac{\sinh(2Bt)}{\sinh(2\Ot t)}x\wedge y} \rbt\text{\hspace{0.5cm}with }
\Ot=\frac{2\Omega}{\theta}
\end{align}
The Gross-Neveu model or the critical Langmann-Szabo-Zarembo models correspond to 
the case $B=\Ot$.
In these models there is no longer any confining decay for the ``long variables''
but only an oscillation:
\begin{align}
Q^{-1}=H^{-1}&=\frac{\Ot}{8\pi}\int_0^\infty \frac{dt}{\sinh(2\Ot t)}\, \exp\lbt -\frac{\Ot}{2}\coth(2\Ot t)(x-y)^2 +
{2\imath\Ot x\wedge y}\rbt\label{eq:CriticalPropa}
\end{align}

This kind of models are called critical. Their construction is more difficult, since
sufficiently many oscillations must be proven independent before power counting
can be established. The prototype paper which solved this problem is \cite{RenNCGN05},
which we briefly summarize now.

The main technical difficulty of the critical models is the absence of decreasing functions for the long $v$ 
variables in the propagator replaced by an oscillation, see (\ref{eq:CriticalPropa}). Note that these 
decreasing functions are in principle created by integration over the $u$ variables\footnote{In all the 
following we restrict ourselves to the dimension $2$.}:
\begin{align}
  \int du\,e^{-\frac{\Ot}{2}\coth(2\Ot t)u^{2}+\imath u\wed v}=&K\tanh(2\Ot t)\,e^{-k\tanh(2\Ot t)v^{2}}.
\end{align}
But to perform all these Gaussian integrations for a general graph is a difficult task (see \cite{RivTan}) 
and is in fact not necessary for a BPHZ theorem. We can instead exploit the vertices and propagators 
oscillations to get rationnal decreasing functions in some linear combinations of the long $v$ variables. 
The difficulty is then to prove that all these linear combinations are independant and hence allow to 
integrate over all the $v$ variables. To solve this problem we need the exact expression of the total 
oscillation in terms of the short and long variables. This consists in a generalization of the Filk's work 
\cite{Filk1996dm}. This has been done in \cite{RenNCGN05}. Once the oscillations are proven 
independant, one can just use the same arguments than in the $\Phi^{4}$ case (see section \ref
{sec:routing-filk-moves}) to compute an upper bound for the power counting:
\begin{lemma}[Power counting $\GN$]\label{lem:compt-puissGN}
  Let $G$ a connected orientable graph. For all $\Omega\in\lsb 0,1\right)$, there exists $K\in\R_{+}$ such 
that its amputated amplitude $A_{G}$ integrated over test functions is bounded by
  \begin{align}
    \labs A_{G}\rabs\les&K^{n}M^{-\frac 12\omega(G)}\label{eq:compt-bound}\\
    \text{with } \omega(G)=&
    \begin{cases}
      N-4&\text{if ($N=2$ or $N\ges 6$) and $g=0$,}\\
      &\text{if $N=4$, $g=0$ and $B=1$,}\\
      &\text{if $G$ is critical,}\\
      N&\text{if $N=4$, $g=0$, $B=2$ and $G$ non-critical,}\\
      N+4&\text{if $g\ges 1$.}
    \end{cases}
  \end{align}
\end{lemma}
As in the non-commutative $\Phi^{4}$ case, only the planar graphs are divergent. But the behaviour of 
the graphs with more than one broken face is different. Note that we already discussed such a feature in 
the matrix basis (see section \ref{sec:prop-et-renorm}). In the multiscale framework, the Feynamn 
diagrams are endowed with a scale attribution which gives each line a scale index. The only subgraphs 
we meet in this setting have all their internal scales higher than their external ones. Then a subgraph $G
$ of scale $i$ is called \emph{critical} if it has $N=4, g=0, B=2$ and that the two ``external'' points in the 
second broken face are only linked by a single line of scale $j<i$. The typical example is the graph of 
figure \ref{fig:sunseti}. In this case, the subgrah is logarithmically divergent whereas it is convergent in 
the $\Phi^{4}$ model. Let us now show roughly how it happens in the case of figure \ref{fig:sunseti} but 
now in $x$-space.

The same arguments than in the $\Phi^{4}$ model prove that the integrations over the internal points of 
the graph \ref{fig:sunseti} lead to a logarithmical divergence which means that $A_{G^{i}}\simeq\cO(1)$ 
in the multiscale framework. But remind that there is a remaining oscillation between a long variable of 
this graph and the external points in the second broken face of the form $v\wed(x-y)$. But $v$ is of order 
$M^{i}$ which leads to a decreasing function implementing $x-y$ of order $M^{-i}$. If these points are 
true external ones, they are integrated over test functions of norm $1$. Then thanks to the additional 
decreasing function for $x-y$ we gain a factor $M^{-2i}$ which makes the graph convergent. But if $x$ 
and $y$ are linked by a single line of scale $j<i$ (as in figure \ref{fig:sunsetj}), instead of test functions 
we have a propagator between $x$ and $y$. This one behaves like (see \eqref{eq:CriticalPropa}):
\begin{align}
  C^{j}(x,y)\simeq&M^{j}\,e^{-M^{2j}(x-y)^{2}+\imath x\wed y}.  
\end{align}
The integration over $x-y$ instead of giving $M^{-2j}$ gives $M^{-2i}$ thanks to the oscillation $v\wed(x-
y)$. Then we have gained a good factor $M^{-2(i-j)}$. But the oscillation in the propagator $x\wed y$ 
now gives $x+y\simeq M^{2i}$ instead of $M^{2j}$ and the integration over $x+y$ cancels the 
preceeding gain. The critical component of figure \ref{fig:sunseti} is logarithmically divergent.

This kind of argument can be repeated and refined for more general graphs to prove that this problem 
appears only when the extrernal points of the auxiliary broken faces are linked only by a \emph{single} 
lower line \cite{RenNCGN05}. This phenomenon can be seen as a mixing between scales. Indeed the 
power counting of a given subgraph now depends on the graphs at lower scales. This was not the case 
in the commutative realm. Fortunately this mixing doesn't prevent renormalisation. Note that whereas the 
critical subgraphs are not renormalisable by a vertex-like counterterm, they are regularised by the 
renormalisation of the two-point function at scale $j$. The proof of this point relies heavily on the fact that 
there is only one line of lower scale.

Let us conclude this section by mentionning the flows of the critical models. One very interesting feature 
of the non-commutative $\Phi^{4}$ model is the boundedness of its flows and even the vanishing of its 
beta function for a special value of its bare parameters \cite
{GrWu04-2,DisertoriRivasseau2006,beta2-06}. Note that its commutative counterpart (the usual $\phi^
{4}$ model on $\R^{4}$) is asymptotically free in the infrared and has then an unbounded flow. It turns 
out that the flow of the critical models are not regularized by the non-commutativity. The one-loop 
computation of the beta functions of the \encv{} Gross-Neveu model \cite{betaGN1Loop} shows that it is 
asymptotically free in the ultraviolet region as in the commutative case. 

\subsection{Non-commutative hyperbolic polynomials}
\label{hyperbo}

Since the Mehler kernel is quadratic it is possible to explicitly compute
the non-commutative analogues of topological or ``Symanzik'' polynomials.

In ordinary commutative field theory, Symanzik's polynomials are obtained
after integration over internal position variables. The amplitude of an 
amputated graph $G$ with external momenta $p$ is, up to a normalization,
in space-time dimension $D$:
\begin{align}
A_G (p) =& \delta(\sum p)\int_0^{\infty} 
\frac{e^{- V_G(p,\alpha)/U_G (\alpha) }}{U_G (\alpha)^{D/2}} 
\prod_l  ( e^{-m^2 \alpha_l} d\alpha_l )\ .\label{symanzik} 
\end{align}
The first and second Symanzik polynomials $U_G$ and $V_G$ are
\begin{subequations}
  \begin{align}
    U_G =& \sum_T \prod_{l \not \in T} \alpha_l \ ,\label{symanzik1}\\
    V_G =& \sum_{T_2} \prod_{l \not \in T_2} \alpha_l  (\sum_{i \in E(T_2)} p_i)^2 \ , \label{symanzik2}
  \end{align}
\end{subequations}
where the first sum is over spanning trees $T$ of $G$
and the second sum  is over two trees $T_2$, i.e. forests separating the graph
in exactly two connected components $E(T_2)$ and $F(T_2)$; the corresponding
Euclidean invariant $ (\sum_{i \in E(T_2)} p_i)^2$ is, by momentum conservation, also
equal to $ (\sum_{i \in F(T_2)} p_i)^2$.

Since the Mehler kernel is still quadratic in position space it is possible
to also integrate explicitly all positions to reduce Feynman amplitudes
of e.g. non-commutative $\phi^4_4$ purely to parametric formulas, but of course
the analogs of Symanzik polynomials are now hyperbolic polynomials which encode 
the richer information about ribbon graphs. The reference for these polynomials
is \cite{gurauhypersyman}, which treats the ordinary $\phi^4_4$ case. In \cite{RivTan}, 
these polynomials are also computed in the more complicated case of critical models.

Defining the antisymmetric matrix $\sigma$ as
\begin{align}
\sigma=&\begin{pmatrix} \sigma_2 & 0 \\ 0 & \sigma_2 \end{pmatrix} \mbox{ with}\\
\sigma_2=&\begin{pmatrix} 0 & -i \\ i & 0 \end{pmatrix}
\end{align}
\noi
the $\delta-$functions appearing in the vertex contribution can be
rewritten as an integral over some new variables $p_V$. We refer to these
variables as to {\it hypermomenta}. Note that one associates such
a hypermomenta $p_V$ to any vertex $V$ {\it via} the relation
\begin{align}
\label{pbar1}
\delta(x_1^V -x_2^V+x_3^V-x_4^V ) =& \int  \frac{d p'_V}{(2 \pi)^4}
e^{ip'_V(x_1^V-x_2^V+x_3^V-x_4^V)}\nonumber\\
=&\int  \frac{d p_V}{(2 \pi)^4}
e^{p_V \sigma (x_1^V-x_2^V+x_3^V-x_4^V)} \ .
\end{align}
\noi

Consider a particular ribbon graph $G$.
Specializing to dimension 4 and choosing a particular root vertex $\bar{V}$ of the graph,
one can write the Feynman amplitude for $G$ in the condensed way
\begin{align}
\label{a-condens}
{\cal A}_G =& \int \prod_{\ell}\big[\frac{1-t_{\ell}^2}{t_{\ell}}\big]^{2} d\alpha_{\ell} \int d x d p e^{-\frac
{\Omega}{2} X G X^t}
\end{align}
where $t_\ell = \tanh\frac{\alpha_\ell}{2}$, 
$X$ summarizes all positions and hyermomenta and $G$ is a certain quadratic form. If we call 
$x_e$ and $p_{\bar{V}}$ the external variables we can decompose $G$ according to an internal 
quadratic form $Q$, an external one $M$ and a coupling part $P$ so that 
\begin{align}
\label{defX}
X =& \begin{pmatrix}
x_e & p_{\bar{V}} & u & v & p\\
\end{pmatrix} \ \ , \ \  G= \begin{pmatrix} M & P \\ P^{t} & Q \\
\end{pmatrix}\ ,
\end{align}
Performing the gaussian integration over all internal variables one obtains:
\begin{align}
\label{aQ}
{\cal A}_G  =& \int \big[\frac{1-t^2}{t}\big]^{2} d\alpha\frac{1}{\sqrt{\det Q}}
e^{-\frac{\Ot}{2}
\begin{pmatrix} x_e & \bar{p} \\
\end{pmatrix} [M-P Q^{-1}P^{t}]
\begin{pmatrix} x_e \\ \bar{p} \\
\end{pmatrix}}\ .
\end{align}
\noi
This form allows to define the polynomials $HU_{G, \bar{v}}$ and $HV_{G, \bar{v}}$, analogs 
of the Symanzik polynomials $U$ and $V$ of the commutative case (see \eqref{symanzik}). 
They are defined by
\begin{align}
\label{polv}
{\cal A}_{{\bar V}}  (\{x_e\},\;  p_{\bar v}) =& K'  \int_{0}^{\infty} \prod_l  [ d\alpha_l (1-t_l^2)^{2} ]
HU_{G, \bar{v}} ( t )^{-2}   
e^{-  \frac {HV_{G, \bar{v}} ( t , x_e , p_{\bar v})}{HU_{G, \bar{v}} ( t )}}.
\end{align}
\noi
They are polynomials in the set of variables $t_\ell$ ($\ell =1,\ldots, L$), the hyperbolic tangent of the
half-angle of the parameters $\alpha_\ell$.

Using now \eqref{aQ} and \eqref{polv} the polynomial $HU_{G,\bar{v}}$ writes
\begin{align}
\label{hugvq}
HU_{\bar{v}}=&(\det Q)^\frac14 \prod_{\ell=1}^L t_\ell
\end{align}

The main results (\cite{gurauhypersyman}) are

\begin{itemize}
 \item The polynomials $HU_{G, \bar{v}}$ and $HV_{G, \bar{v}}$ 
have a strong positivity property. Roughly speaking they are sums of monomials
with positive integer coefficients. This positive integer property 
comes from the fact that each such coefficient is the square of a Pfaffian 
with integer entries,

\item {Leading terms} can be identified in a given ``Hepp sector'', 
at least for \textit{orientable graphs}.
A Hepp sector is a complete ordering of the $t$ parameters.
These leading terms which can be shown strictly positive in $HU_{G, \bar{v}}$ correspond to super-trees
which are the disjoint union of a tree in the direct graph and a tree in the dual graph.
Hypertrees in a graph with $n$ vertices and $F$ faces have therefore $n+F-2$ lines.
(Any connected graph has hypertrees, and under reduction of the hypertree, the graph becomes
a hyperrosette). Similarly one can identify ``super-two-trees'' $HV_{G, \bar{v}}$ 
which govern the leading behavior of $HV_{G, \bar{v}}$ in any Hepp sector.
\end{itemize}

From the second property, one can deduce the \textit{exact power counting} of any orientable
ribbon graph of the theory, just as in the matrix base.

Let us now borrow from \cite{gurauhypersyman} some examples of these hyperbolic polynomials. 
We put $s =(4\theta\Omega)^{-1}$.
\begin{figure}[htb] 
  \centering
  \includegraphics{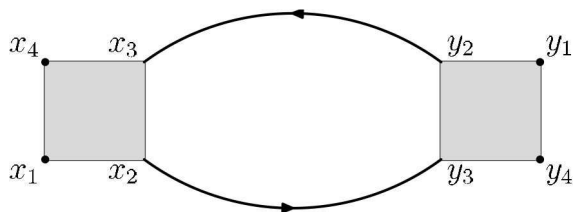}
  \caption{The bubble graph}
  \label{figex1}
\end{figure}
For the bubble graph of figure \ref{figex1}:
\begin{align}
HU_{G,v}=&(1+4s^2)(t_1+t_2+t_1^2t_2+t_1t_2^2)\,,\nonumber\\
HV_{G,v}=&t_2^2\Big{[}p_2+2s(x_4-x_1)\Big{]}^2+t_1t_2\Big{[}2p_2^2+(1+16s^4)(x_1-x_4)^2
                \Big{]}\,,\nonumber\\
                &+t_1^2\Big{[}p_2+2s(x_1-x_4)\Big{]}^2\nonumber\\
\end{align}

For the sunshine graph fig.~\ref{figex2}:
\begin{figure}[htb] 
  \centering
  \includegraphics{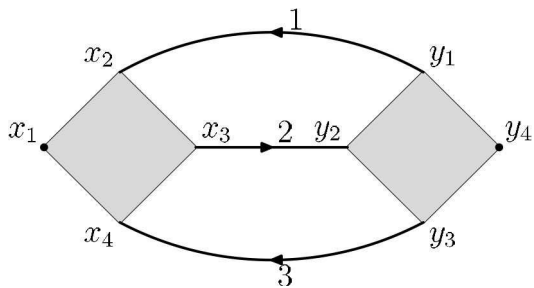}
  \caption{The Sunshine graph}
  \label{figex2}
\end{figure}
\begin{align}
  HU_{G,v}=&\Big{[} t_1t_2+t_1t_3+t_2t_3+t_1^2t_2t_3+t_1t_2^2t_3+t_1t_2t_3^2\Big{]}
  (1+8s^2+16s^4)\nonumber\\
  &+16s^2(t_2^2+t_1^2t_3^2)\, ,\nonumber\\
\end{align}

For the non-planar sunshine graph (see fig.~\ref{figex3}) we have:
\begin{figure}[htb] 
  \centering
  \includegraphics{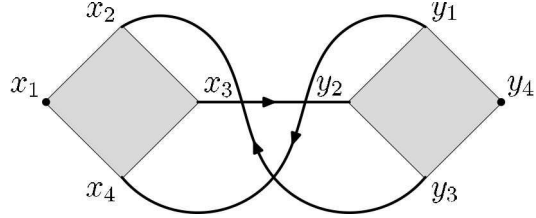}
  \caption{The non-planar sunshine graph}
  \label{figex3}
\end{figure}
\begin{align}
  HU_{G,v}=&\Big{[} t_1t_2+t_1t_3+t_2t_3+t_1^2t_2t_3+t_1t_2^2t_3+t_1t_2t_3^2\Big{]}
  (1+8s^2+16s^4)\nonumber\\
  &+4s^2\Big{[}1+t_1^2+t_2^2+t_1^2t_2^2+t_3^2+t_1^2t_3^2+t_2^2t_3^2+
  t_1^2t_2^2t_3^2\Big{]}\,,\nonumber
\end{align}
We note the improvement in the genus with respect to its planar counterparts.

For the broken bubble graph (see fig. \ref{figex4}) we have:
\begin{figure}[htb] 
  \centering
  \includegraphics{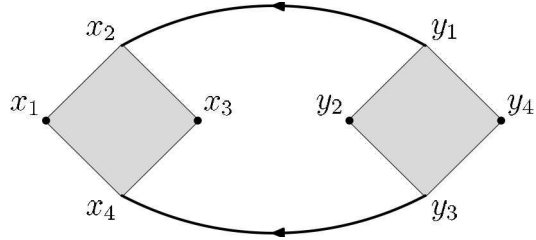}
  \caption{The broken bubble graph}
  \label{figex4}
\end{figure}
\begin{align}
  HU_{G,v}=&(1+4s^2)(t_1+t_2+t_1^2t_2+t_1t_2^2)\,,\nonumber\\
  HV_{G,v}=& t_2^2 \Big{[}4s^2(x_1+y_2)^2+(p_2-2s(x_3+y_4))^2\Big{]}+t_1^2\Big{[}p_2
  +2s(x_3-y_4) \Big{]}^2\,,\nonumber\\
  &+t_1t_2\Big{[}8s^2y_2^2+2(p_2-2sy_4)^2+(x_1+x_3)^2+16s^4(x_1-x_3)^2\Big{]}\nonumber\\
  &+t_1^2t_2^24s^2(x_1-y_2)^2\,,\nonumber
\end{align}
Note that $HU_{G,v}$ is identical to the one of the bubble with only one broken face. 
The power counting improvement comes from the broken face and can be seen only in $HV_{G,v}$. 

\begin{figure}[htb] 
  \centering
  \includegraphics{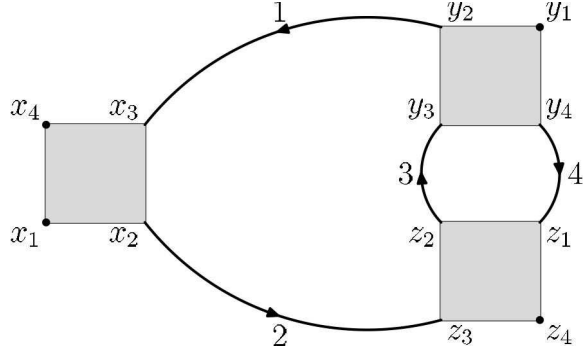}
  \caption{The half-eye graph}
  \label{figeye}
\end{figure}
Finally, for the half-eye graph (see Fig. \ref{figeye}), we start by defining:
\begin{align}
  A_{24}=&t_1t_3+t_1t_3t_2^2+t_1t_3t_4^2+t_1t_3t_2^2t_4^2\,.
\end{align}
The $HU_{G,v}$ polynomial with fixed hypermomentum corresponding to the vertex with two external 
legs is:
\begin{align}\label{hueye1}
  HU_{G,v_1}=&(A_{24}+A_{14}+A_{23}+A_{13}+A_{12})(1+8s^2+16s^4)\nonumber\\
  &+t_1t_2t_3t_4(8+16s^2+256s^4)+4t_1t_2t_3^2+4t_1t_2t_4^2\nonumber\\
  &+16s^2(t_3^2+t_2^2t_4^2+t_1^2t_4^2+t_1^2t_2^2t_3^2)\nonumber\\
  &+64s^4(t_1t_2t_3^2+t_1t_2t_4^2)\,,
\end{align}
whereas with another fixed hypermomentum we get:
\begin{align}
HU_{G,v_2}=&(A_{24}+A_{14}+A_{23}+A_{13}+A_{12})(1+8s^2+16s^4)\nonumber\\
&+t_1t_2t_3t_4(4+32s^2+64s^4)+32s^2t_1t_2t_3^2+32s^2t_1t_2t_4^2\nonumber\\
&+16s^2(t_3^2+t_1^2t_4^2+t_2^2t_4^2+t_1^2t_2^3t_3^2)\,.\label{hueye2}
\end{align}

Note that the leading terms are identical and the choice of the root perturbs only the non-leading ones. 
Moreover note the presence of the $t_3^2$ term. Its presence can be understood by the fact that in the 
sector $t_1,t_2,t_4>t_3$ the subgraph formed by the 
lines $1,2,4$ has two broken faces. This is the sign of 
a power counting improvement due to the additional broken face in that sector. 
To exploit it, we have just to integrate over the variables of line $3$
in that sector, using the second polynomial $HV_{G',v}$ for the triangle subgraph $G'$ 
made of lines $1,2,4$.  

In the \textit{critical case}, it is essential to introduce arrows upon the lines
and to take them into account. The corresponding analysis together with many examples 
are given in \cite{RivTan}.

\subsection{Conclusion}

Non-commutative QFT seemed initially to have non-renormalisable divergencies,
due to UV/IR mixing. But following the Grosse-Wulkenhaar breakthrough, 
there has been recent rapid progress in our understanding of renormalisable QFT on Moyal spaces.
We can already propose a preliminary classification of these models into 
different categories, according to the behavior of their propagators:
\begin{itemize}
\item ordinary models at $0 < \Omega < 1$ such as $\phi^4_4$ 
(which has non-orientable graphs) or $(\bar\phi\phi)^2$ models 
(which has none). Their
propagator, roughly $(p^2 + \Omega^2 \tilde x^2 + A)^{-1}$ is LS covariant and has good decay both in 
matrix space (\ref{th1}-\ref{thsummax}) and direct space (\ref{tanhyp}). They have non-logarithmic mass 
divergencies and definitely require ``vulcanization'' i.e. the $\Omega$ term.
\item ``supermodels'', namely ordinary models but at $\Omega = 1$ in which the propagator is LS 
invariant.
Their propagator is even better. In the matrix base 
it is diagonal, e.g. of the form $G_{m,n}= (m+n + A)^{-1}$, where
$A$ is a constant. The supermodels seem generically 
ultraviolet fixed points of the ordinary models, at which non-trivial Ward identities
force the vanishing of the beta function. The flow of $\Omega$ to the $\Omega = 1$ fixed
point is very fast (exponentially fast in RG steps).
\item ``critical models'' such as orientable versions of LSZ or Gross-Neveu (and presumably 
orientable gauge theories
of various kind: Yang-Mills, Chern-Simons...). They may have only logarithmic divergencies and 
apparently no perturbative UV/IR mixing. However the vulcanized version 
still appears the most generic framework for their treatment.
The propagator is then roughly $(p^2 + \Omega^2 \tilde x^2 + 2\Omega \tilde x \wedge p)^{-1}$.
In matrix space this propagator shows definitely a weaker decay (\ref{mainbound1})
than for the ordinary models, because of the presence of a non-trivial saddle point.
In direct space the propagator no longer decays with respect to the long variables, 
but only oscillates. Nevertheless the main lesson is that 
in matrix space the weaker decay can still be used; and in
$x$ space the oscillations can never be completely killed by the vertices
oscillations. Hence these models retain therefore essentially the power counting 
of the ordinary models, up to some nasty details concerning the
four-point subgraphs with two external faces.
Ultimately, thanks to a little conspiration in which the
four-point subgraphs with two external faces are renormalised by the mass renormalisation,
the critical models remain renormalisable. This is the main message of \cite{RenNCGN05,vignes-tourneret06:PhD}.
\item ``hypercritical models'' which are of the previous type but at $\Omega = 1$. Their propagator in the 
matrix base is diagonal and depends only on one index $m$
(e.g. always the left side of the ribbon). It is of the form $G_{m,n}= (m + A)^{-1}$.
In $x$ space the propagator oscillates in a way that often
exactly compensates the vertices oscillations. These models have definitely
worse power counting than in the ordinary case, with e.g. quadratically divergent 
four point-graphs (if sharp cut-offs are used). Nevertheless Ward identities
can presumably still be used to show that they can still be renormalised. This 
probably requires a much larger conspiration to generalize
the Ward identities of the supermodels.
\end{itemize}
Notice that the status of non-orientable critical theories is not yet clarified.

Parametric representation can be derived in the non-commutative case. It implies 
hyper-analogs of Symanzik polynomials which condense the 
information about the rich topological structure of a ribbon graph.
Using this representation, dimensional regularization and dimensional renormalisation 
should extend to the non-commutative framework.

Remark that trees, which are the building blocks of the Symanzik polynomials, are also 
at the heart of (commutative) constructive theory, whose philosophy
could be roughly summarized as ``You shall use trees\footnote{These trees
may be either true trees of the graphs in the Fermionic case or trees associated to 
cluster or Mayer expansions in the Bosonic case, but this distinction is not essential.}, 
but you shall \textit{not} develop their loops or else you shall diverge''.
It is quite natural to conjecture that hypertrees, which are the natural non-commutative
objects intrinsic to a ribbon graph, should play a key combinatoric role
in the yet to develop non-commutative constructive field theory.

In conclusion we have barely started to scratch the world of 
renormalisable QFT on non-commutative spaces.
The little we see through the narrow window now open
is extremely tantalizing. There {exists renormalizable NCQFTs} eg $\phi^4$ on ${\mathbb R}^4_\theta$, 
Gross-Neveu on ${\mathbb R}^2_\theta$ and they 
seem to enjoy better propoerties than their commutative counterparts,
for instance they no longer have Landau ghosts! 
Non-commutative non relativistic field theories with a chemical potential
seem the right formalism for a study {ab initio} of condensed matter in 
presence of a magnetic field, and in particular of the Quantum Hall Effect.
The correct scaling and RG theory of this effect presumably requires to build
a very singular theory (of the hypercritical type) because of the huge
degeneracy of the Landau levels. To understand this theory and 
the gauge theories on non-commutative spaces
seem the most obvious challenges ahead of us.


\end{document}